\documentclass[fleqn,10pt]{wlscirep}
\usepackage[utf8]{inputenc}
\usepackage[T1]{fontenc}
\usepackage{subcaption}
\usepackage[T1]{fontenc}
\DeclareUnicodeCharacter{2212}{-}
\title{The amplitudons of dynamical stable binary cubic crystals 
       for hot-electron cooling in BaTe, CdTe, SrTe and SnTe  
       for enhanced photo voltaic effect.}

\author[1,*]{T.E.Ada}
\author[2]{L.D.~Deja}
\author[3]{D.A.~Adem}
\author[2,*]{K.N.~Nigussa}
\affil[1]{Department of Physics,\ Dilla University,\ P.O. Box 419,\ 
         Dilla,\ Ethiopia}
\affil[2,3]{Department of Physics,\ Addis Ababa University,\ P.O. Box 1176,\ 
         Addis Ababa,\ Ethiopia}
\affil[3]{Department of Physics,\ Wolkite University,\ P.O. Box 07,\ 
         Wolkite,\ Ethiopia}

\affil[*]{corresponding.kenate.nemera@aau.edu.et}
\affil[*]{corresponding.tewodros.eyob@aau.edu.et}
\begin{abstract}
Dynamical properties of materials determined using force constant expansions in the HIPHIVE package and density functional theory as implemented in the Gpaw code. Thermal conductivity of alkaline earth tellurides was determined, and the narrower width between amplitudons and electrical modes exhibited supersonic cooling and heat transportation, and our findings are in perfect agreement with experimental research. When compared to the other binary crystals, BaTe appears to offer the dual advantage of prolonged photon lifetime at normal temperature as well as a speedier thermo-electric converter due to the tiny gap between amplitudons and electric modes.
\end{abstract}
\begin{document}

\flushbottom
\maketitle
\thispagestyle{empty}

%\noindent Please note: Abbreviations should be introduced at the first mention in the main text – no abbreviations lists. Suggested structure of main text (not enforced) is provided below.

\section*{Introduction}
Thermoelectric materials have attracted significant research in recent years due to their promise in energy applications such as converting waste heat into power and replacing mechanical cooling systems with more environmentally friendly thermoelectric devices. For decades. in fact, the goal has been to improve the energy conversion efficiency, which is characterized by the dimensionless figure of merit, zT, which is defined as the ratio of the electronic power factor and the thermal conductivity of the thermoelectric material, and fixed to only $zT=1.0$. However, a surprising 40$\%$ increase in zT value from 1.0 to 1.4 was accomplished by using a ball milling and hot pressing procedure on an ingot~\cite{Poudel2008, CWood1988}. A further 50$\%$ improvement has been made to that of $\mathrm{Bi_{2}Te_{3}}$ ingot materials accounting for zT 1.56 at 300 k~\cite{XWTXYYZQT2009}. Because anharmonic lattice dynamics necessitate both high electronic power factors and low thermal conductivities~\cite{NMTT1999}, increasing the figure of merit is to reduce the phonon contribution to thermal transport while maintaining electrical conductivity unaltered, either by a high Seebeck coefficient or a low thermal conductivity~\cite{Delaire2011}. Identifying a phonon blocking mechanism is thus beneficial. 

One example of how defected crystals can affects a material's electrical properties is by scattering or localizing electrons, which in turn decreases the power factor. Another way that lattice disorder can impact a material is by introducing loosely bound rattling atoms into crystals with complex unit cells, which can limit the flow of heat and make thermoelectric conversion less efficient~\cite{Chen2017, Manley2018}. 

It is evident that  the presence of two or more mutually incompatible elements of translational symmetry implies the presence of continuously accumulated phase shifts, causing an acoustic-like excitation coming from the incommensurate modulation consisting of two modes amplitudon (optic-phonon-like) and phason (acoustic phonon-like). The temperature dependence of the wave vector near commensurate phase forms solitons, which are completely analogous to phonons in a regular atomic lattice; however, the soliton lattice is faster than the crystal lattice. Thus, phasons travel faster than acoustic phonons, reduce domain wall energy, and favor incommensurate phases~\cite{BPVB1980, MQRC1983}.

In what is now known as the FPU recurrent phenomena, researchers discovered that energy traveling between degrees of freedom remained unevenly distributed at any scale. Some complicated nonlinear systems can be chaotic or extremely complex while still being predicted to be ergodic. However, some nonlinear systems, such as the FPU simulation, create coherent nonlinear structures and traveling waves.

For instance, a soliton is a solitary traveling wave solution that is the quintessential coherent structure encountered in many macroscopic nonlinear systems. In addition to the previously described efforts on nonlinear traveling waves in one dimensional lattices, Mathew \emph{et al}~\cite{Bryan2020} have reported three-dimensional crystal lattices that allow anharmonicity-driven intrinsic localized modes (ILMs), which are stationary vibrations limited to only a few atoms.

Localization and related changes in the lattice dynamics in a PbSe crystal using inelastic neutron and x-ray scattering, on the other hand, show that localization occurs at close to predicted temperatures but involves more spectral weight than expected and drives unanticipated changes in the lattice dynamics, including an unexpected sharpening of the longitudinal acoustic (LA) phonon at high temperatures. Instead of localizing with a fraction of the intensity of normal phonons~\cite{SNHOMA2017}, the entire spectral weight of a significant portion of the transverse optic phonon immediately acquires flat dispersion (zero group velocity) and appears frequency fragmented.

The anharmonic dynamics transition can be seen in the optic phonon's localization (flattening) and fragmentation~\cite{Manley2011, MJBZA2014}. This transition has also been found as a minor kink in the thermal diffusivity data, which is analogous to ILM ordering in NaI.

The discovery of in-band localization in a PbSe crystal not only broadens the domain of anharmonic localization, but also has crucial implications for the low thermal conductivity required for thermoelectric performance. One can see that the reorganization of spectral characteristics caused by localization profoundly alters the phase space for scattering, which explains the sharpening of the LA phonon. These findings demonstrate that nonlinear physics beyond standard harmonic perturbations can play a significant role in influencing vibrational transport.
%%%%%%%%%%%%Figure-1%%%%%%%%%%%%%%%%%%%%%%%%%%%%%
\begin{figure}[htbp!]
\centering
\includegraphics[scale=0.35]{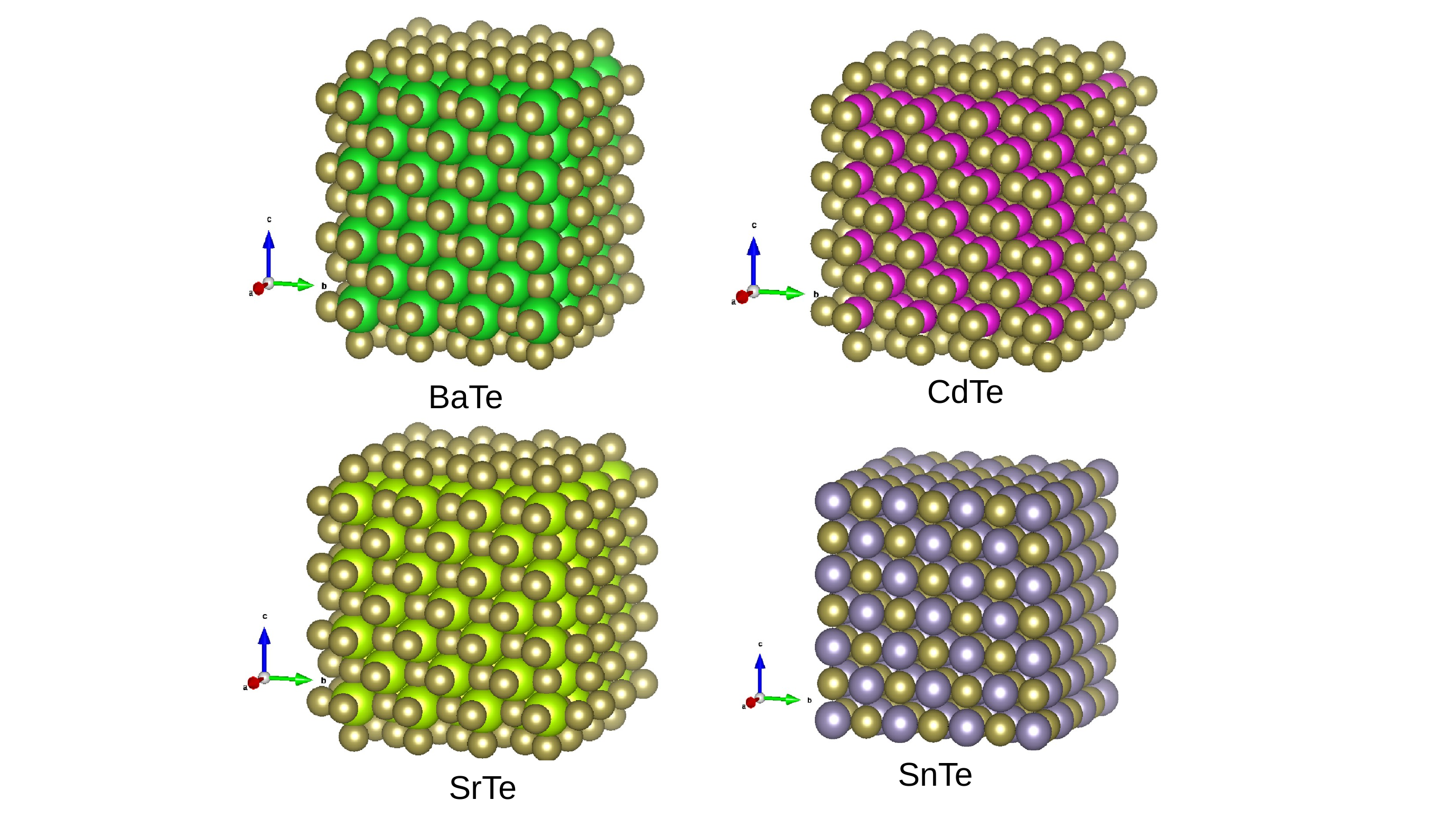}
\caption{ Cubic crystal structures of BaTe, SrTe, and SnTe were substituted on the prototype CdTe system to highlight the incommensurate structure considerations. 
\label{fig1}}
\end{figure}
%%%%%%%%%%%%%%%%%%%%%%%%%%%%%%%%%%%%%%%%%%%%%%%%%%

CdTe has been identified as one of the most promising absorber materials for thin-film solar cells. As a result, continued research activities in CdTe-based photovoltaics remain highly appealing. According to one claim, substituting or doping of CdTe-material optimizes carrier collection~\cite{FCGRC2006, CFK2006, HWK2009}, resulting in a considerable improvement in CdTe solar-cell performance.

Here we look at the incommensurate modulation originates due to Ba, Sr and Sn substituted in CdTe crystal system. CdTe is Halite, Rock Salt structured and crystallizes in the cubic  space group. The corner-sharing octahedral are not tilted~(See Fig.~\ref{fig1}).

\section{Methodology}
In crystals, it is assumed that atoms move around their equilibrium position, thus, the potential surface can be expanded in a Taylor series with respect to the atomic displacements, \textbf{u} relative to a set of reference positions, $R_{0}$.
\begin{equation}
\begin{aligned}
V&=\Phi_{0}+\sum_{i}\sum_{a}\Phi_{i}^{\alpha}u_{i}^{\alpha}+\frac{1}{2!}\sum_{ij}\sum_{\alpha\beta}\Phi_{ij}^{\alpha\beta}u_{i}^{\alpha}u_{j}^{\beta}\\
&+\frac{1}{3!}\sum_{ijk}\sum_{\alpha\beta\gamma}\Phi_{ijk}^{\alpha\beta\gamma}u_{i}^{\alpha}u_{j}^{\beta}u_{k}^{\gamma}+\frac{1}{4!}\sum_{ijkl}\sum_{\alpha\beta\gamma\eta}\Phi_{ijkl}^{\alpha\beta\gamma\eta}u_{i}^{\alpha}u_{j}^{\beta}u_{k}^{\gamma}u_{l}^{\eta}+\cdots
\label{eq1}
\end{aligned}
\end{equation}
Where $\Phi$ is crystal potential, while Latin and Greek indices enumerate atoms and Cartesian coordinates, respectively. From eq.~\ref{eq1} the coefficients of series expansions, $\Phi_{0}$, $\Phi_{i}^{\alpha}$, $\Phi_{ij}^{\alpha\beta}$, and $\Phi_{ijk}^{\alpha\beta\gamma}$ are the zeroth, first, second, and third order force constants, respectively. Atomic vibration is solved with second order terms as harmonic approximation with small displacement at constant volume. Thus, the force acting on atom $\textit{i}$ along direction $\alpha$ is given by
\begin{equation}
\begin{aligned}
F_{i}^{\alpha}=-\frac{\partial V}{\partial u_{i}^{\alpha}}=-\sum_{j}\sum_{\beta}\Phi_{ij}^{\alpha\beta}u_{j}^{\beta}-\frac{1}{2!}\sum_{jk}\sum_{\beta\gamma}\Phi_{ijk}^{\alpha\beta\gamma}u_{j}^{\beta}u_{k}^{\gamma}\\
-\frac{1}{3}\sum_{jkl}\sum_{\beta\gamma\eta}\Phi_{ijkl}^{\alpha\beta\gamma\eta}u_{j}^{\beta}u_{k}^{\gamma}u_{l}^{\eta}+\cdots
\end{aligned}
\end{equation}
Applying derivative operator  to  a force $F_{i}=-\frac{\partial \Phi}{\partial u_{i}}$, one come up with second order force constants,
\begin{equation}
\frac{\partial^{2} \Phi}{\partial u_{\alpha} \partial u_{i}}=-\frac{\partial F_{i}}{\partial u_{\alpha}}
\end{equation}
Dynamical property of atoms in the harmonic approximation is obtained by solving eigenvalue problem of dynamical matrix, $D(q)$.
\begin{equation}
\sum D(q)e_{qj}=\omega^{2}e_{qj}
\end{equation}

Solving Newton's equations of motion requires solving for the eigenvalues of the dynamical matrix, the eigenvalues of the dynamical matrix are the squared frequencies. Note that, each frequency corresponds to a phonon mode, $e_{qj}$ dynamical matrix is a function of the wave vector, $\textit{q}$ with associated $\textit{j}$ dimensional Cartesian axises. Since $D(q)$ is an Hermitian matrix, its eigenvalues, $\omega^{2}_{qj}$, are real. Phononic characteristics could indicate dynamical stability of crystal structure, for instance, phonons with real $\&$ positive frequency indicates that the system is stable, however, imaginary frequency or negative eigenvalue show dynamical instability of a system which means the corrective atomic displacement required or imaginary mode give a clue to study displacve phase transition~\cite{TIT2015}. 
The  phonon density of states  is calculated as
\begin{equation}
g(\omega)=\frac{1}{N}\delta(\omega-\omega_{qj})
\end{equation}
Where N is the number of unit cells in crystal, phonon mode at unique frequency over Brillouin zone plotted as as shown in Fig.~(\ref{fig5}) the band structure explains acoustic modes damped gradually to $\Gamma$-center which relates to lower thermal conductivity, further decrement in thermal conductivity expected due to presence of many phonon modes, the energy of phonon system determined as
\begin{equation}
E=\sum_{qj}\hbar \omega_{qj}\left[\frac{1}{2}+\frac{1}{exp(\hbar\omega_{qj}/k_{B}T)-1}\right]
\end{equation}
Where T, $k_{B}$, and $\hbar$ are the temperature, the Boltzmann constant, and the reduced Plank constant, respectively. Using the thermodynamic relations, a number of thermal properties, such as constant volume heat capacity, $C_{V}$ 
\begin{equation}
C_{V}=\sum_{qj}C_{qj}=\sum_{qj}k_{B}\left(\frac{\hbar\omega_{qj}}{k_{B}T}\right)^{2}\frac{exp(\hbar\omega_{qj}/k_{B}T)}{[exp(\hbar\omega_{qj}/k_{B}T)-1]^{2}},
\end{equation}

Helmholtz free energy F,

\begin{equation}
F=\frac{1}{2}\sum_{qj}\hbar\omega_{qj}+k_{B}T\sum_{qj}\mathrm{ln[1-exp(-\hbar\omega_{qj}/k_{B}T)]},
\end{equation}

and entropy S,
\begin{equation}
\begin{aligned}
S&=\frac{1}{2T}\sum_{qj}\hbar\omega_{qj}coth[\hbar\omega_{qj}/2k_{B}T]\\
&-k_{B}\sum_{qj}\mathrm{ln[2sinh(\hbar\omega_{qj}/2k_{B}T)]}
\end{aligned}
\end{equation}

The diagonal components of the thermal conductivity tensor are where V$_{\alpha qj}$  is the group velocity of the mode, $q_{j}$ and $\tau_{\alpha q}$ is the mode lifetime for transport in direction $\alpha$, $C_{q}$ is the mode heat capacity, and V is volume. As illustrated in Fig.(\ref{fig6}), thermalconductivity is supposed to be insensitive to temperature changes $\&$ computed as,
\begin{equation}
\kappa_{\alpha\alpha}=\frac{1}{V}\sum_{qj}C_{qj}V_{\alpha qj}\tau_{\alpha qj}
\end{equation}

Phonon calculation~\cite{TACLTI2015} explains thermal properties, mechanical properties, phase transition, and superconductivity. Thermodynamic variables at constant volume is transformed to those at constant pressure. 
\begin{equation}
G(T,P)=\mathrm{min_{V}}[F(T; V)+ PV]
\end{equation}
Gibbs free energy $\mathrm{G(T, P)}$ at given temperature, T and pressure, P is obtained from Helmholtz free energy $\mathrm{F(T; V)}$ through the transformation. Phonon lifetime is crucial in search of thermoelectric materials, transistors, and heat management systems since it could explain how long it takes phonon mode vibrations to dissipate~\cite{BJHAFZFEEFATKAKHEP2022}. Peaks in Fig.~(\ref{fig4}) represent key phonon modes that contribute to overall phonon life time throughout frequency, which corresponds to temperature.

Here, we investigate a BaTe, SrTe, SnTe cubic dimmer from the space group Fm-3m (International Tables of Crystallography number 225). Random displacements were applied to perfect supercells with up to 216 atoms (equal to 3$\times$3$\times$3 unit cells) to generate input configurations. The average displacement amplitudes for these configurations were around 0.0098, resulting in average forces of around 230 meV per angstrom and maximum forces of around 0.6 eV per angstrom. DFT computations were used to generate reference forces, which were obtained using the projector augmented wave method~\cite{Blochl94, MHJ2005, KS65} as implemented in GPAW code~\cite{Enkovaaraetal2010} and the vdW-DF-cx method~\cite{VSYJCMFPH2021}, which combines semi-local exchange with nonlocal correlation. The Brillouin zone was sampled with Monkhorst Pack k-point grids, which are equivalent to a 2$\times$2$\times$2 mesh relative to the primitive cell. The plane-wave energy cutoff was set to 333 eV, the forces were evaluated using a finer grid, and the reciprocal projection scheme was used throughout the calculation.

Fig.(\ref{fig5}) depicts the phonon dispersion produced using hiphive~\cite{EFFEEP2019} with all invariant constraints imposed. The lowermost transverse acoustic branch clearly demonstrates quadratic dispersion in this case. If the rotational sum rules are deactivated, the dispersion is almost unchanged except for the appearance of a small imaginary pocket in the immediate vicinity of the Brillouin center, and thus does not yield a quadratic dispersion, but small errors can have a significant effect whenever the acoustic modes contribute significantly to a property, such as thermal conductivity.The thermal conductivity was calculated using the phono3py code in the setting of phonon Boltzmann transport theory in the relaxation time approximation~\cite{CL2013}. The Brillouin integration was performed using the tetrahedron method and a 13$\times$13$\times$13 q-point mesh. Only phonon-phonon scattering was regarded as a rate-limiting process for clarity. Reference computations were performed on pairings with distances up to 6.0  angstrom.

For all systems, a fourth-order FCP was generated with cutoff radii of 6.0, 5.0, and 5.0 angstrom for second, third, and fourth-order terms, respectively, corresponding to 216 clusters in the unit cell. The associated cluster space included 34 distinct orbits (6, 8, and 20 for second-, third-, and fourth-order, respectively), with interactions up to the fourth nearest neighbor for the pairings, resulting in 194 free parameters.

Training data consisted of five structures with 216 atoms (3$\times$3$\times$3 conventional unit cells) for a total of 3240 force components. The structures were created by using displacements taken at random from a normal distribution and tweaked to avoid inter-atomic distances shorter than 2.3 angstrom. The average atomic displacement as a result was around 0.13 angstrom. The reference forces were calculated using a grid-based augmented plane wave implemented in the Gpaw code. Because the system is highly over determined, the model parameters were trained using traditional least-squares fitting.

\section{Results and discussion}
The optical photoabsorption spectrum and nonlinear effects can be estimated using time-dependent external perturbation, and interactions inside a material can be investigated by linearly combining atomic orbitals. As demonstrated in Fig.~(\ref{fig4}), the number of phonon modes at 0.5THz frequency is substantially more than the number of modes in CdTe, SrTe, and SnTe, respectively.

%%%%%%%%%%%%%%%%%%%%%%%%%%%%Figure-2%%%%%%%%%%%%%%%%%%%%%%%%%%%%%
\begin{figure}[htbp!]
        \centering
        \includegraphics[scale=0.35]{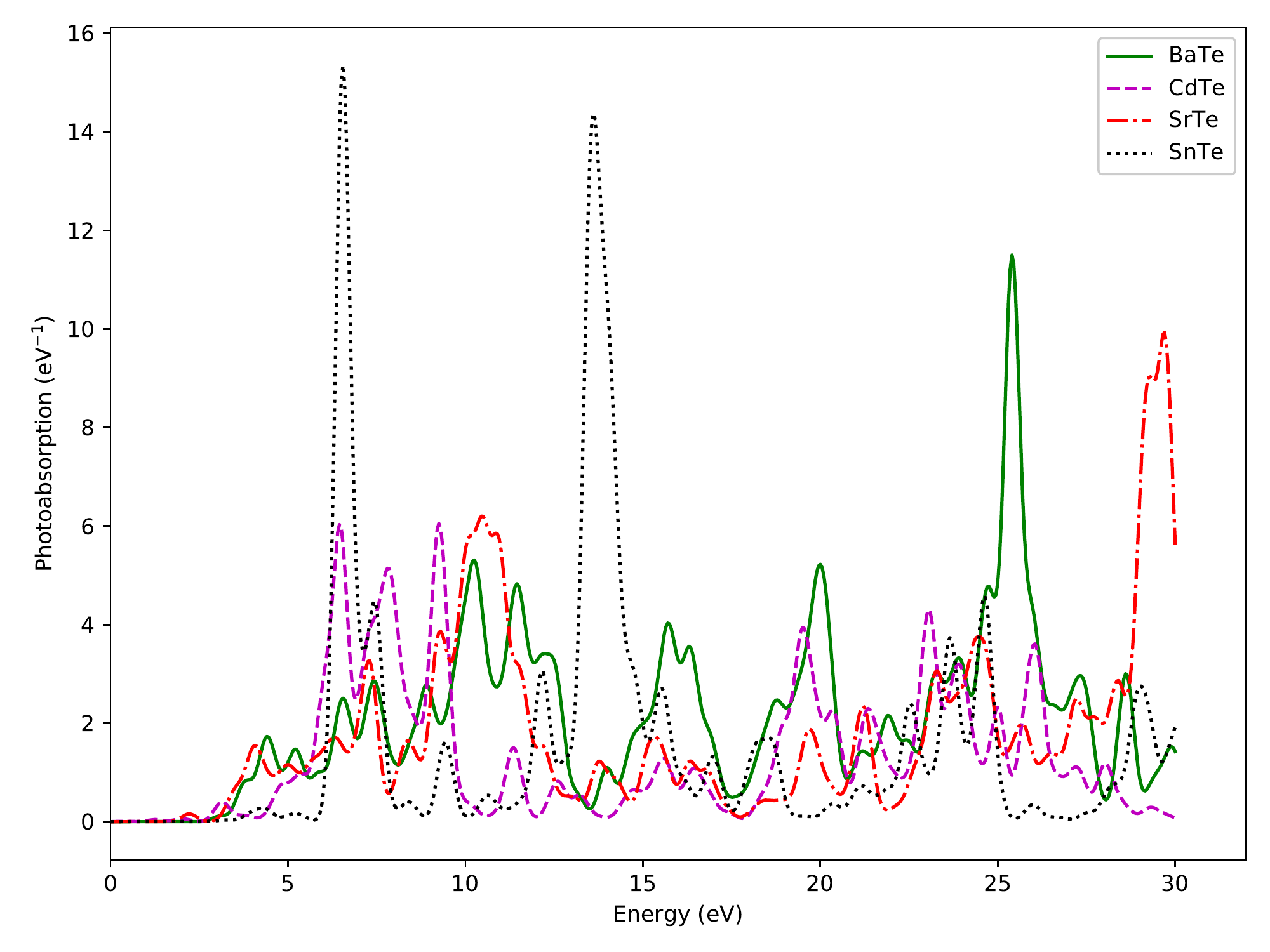}
        \caption{\label{fig2} The photo absorption spectrum of
         the $\mathrm{\{Ba, Cd, Sr, \&~ Sn\}}$Te systems along z-axis
         for 30fs of time step and at kick strength of $1e^{-3}$
         with grid parameter, $\mathrm{h=0.3}$, vacuum set to 6.0 with
          LCAO basis set assumed for constituent atom.}
\end{figure}
\begin{figure}[htpb!]
        \centering
        \includegraphics[scale=0.35]{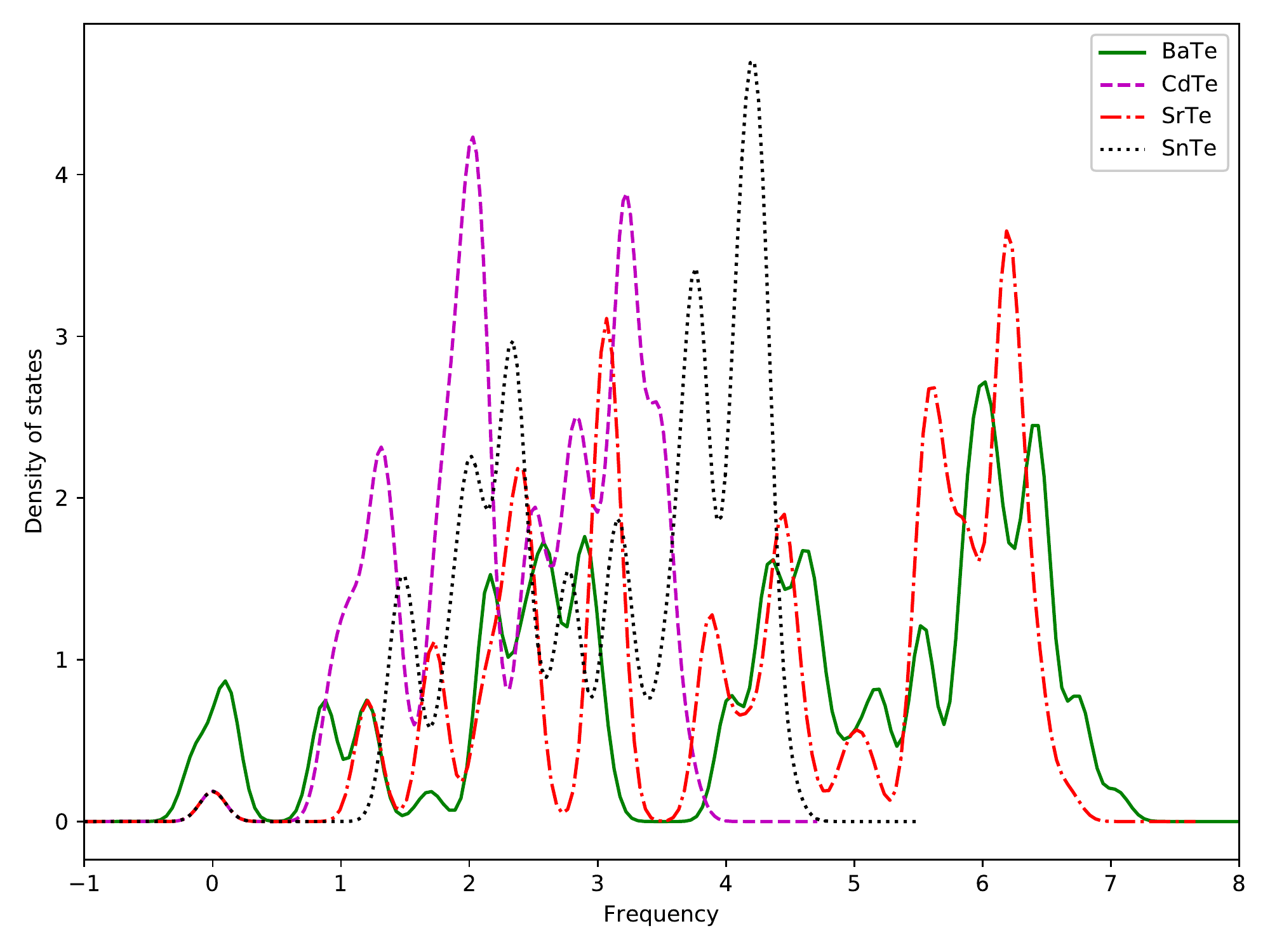}
        \caption{\label{fig3} Density of states that contribute most to
        the lattice vibration of $\mathrm{\{Ba, Cd, Sr, \&~ Sn\}}$Te systems.
        For the interpretations of the references to color in this
        plot legend, the reader is referred to the web-version.}
\end{figure}
\begin{figure*}[htbp!]
\begin{subfigure}[t]{0.5\textwidth}
\centering
\mbox{\includegraphics[scale=0.35]{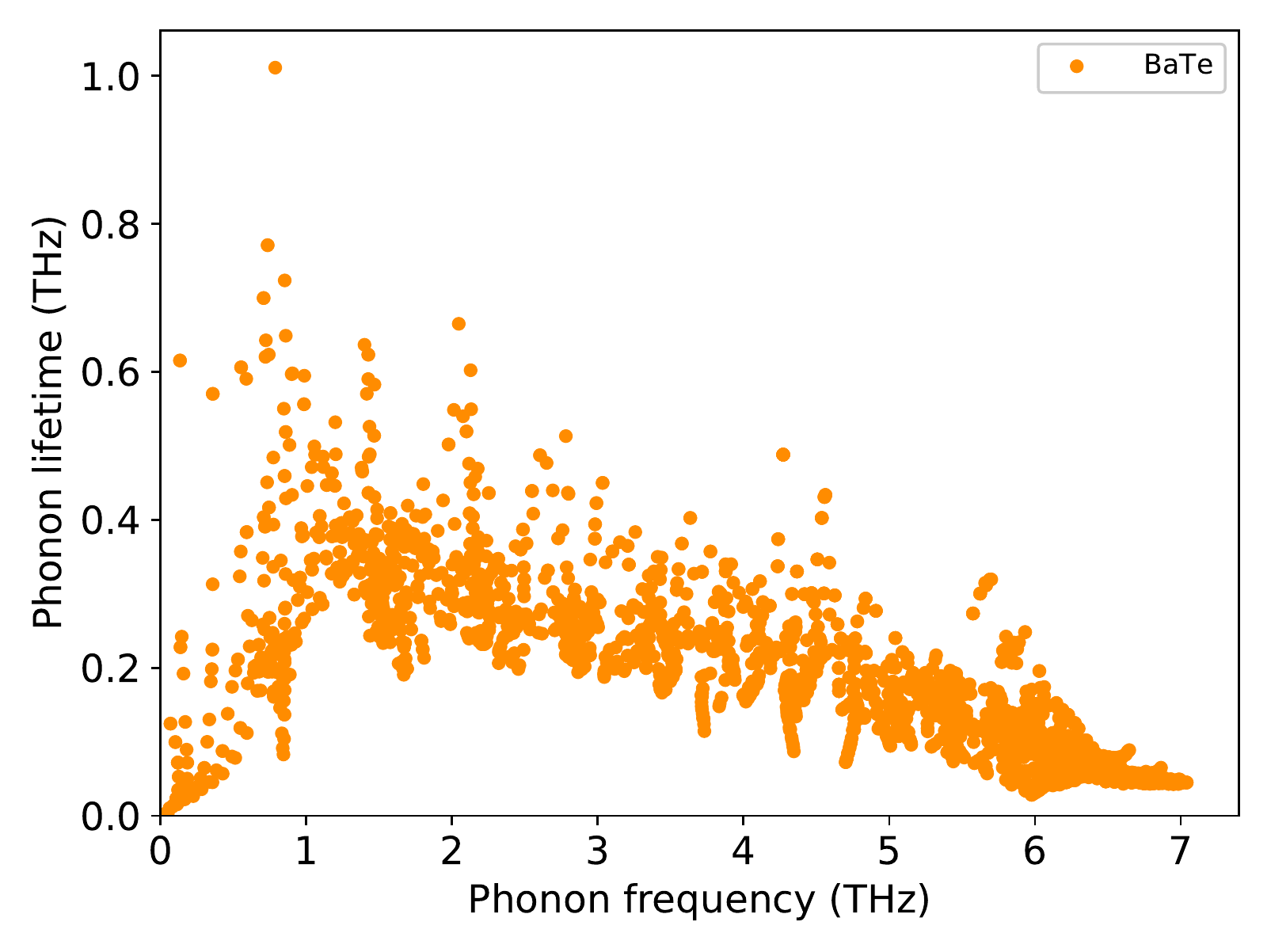}}
\mbox{\includegraphics[scale=0.35]{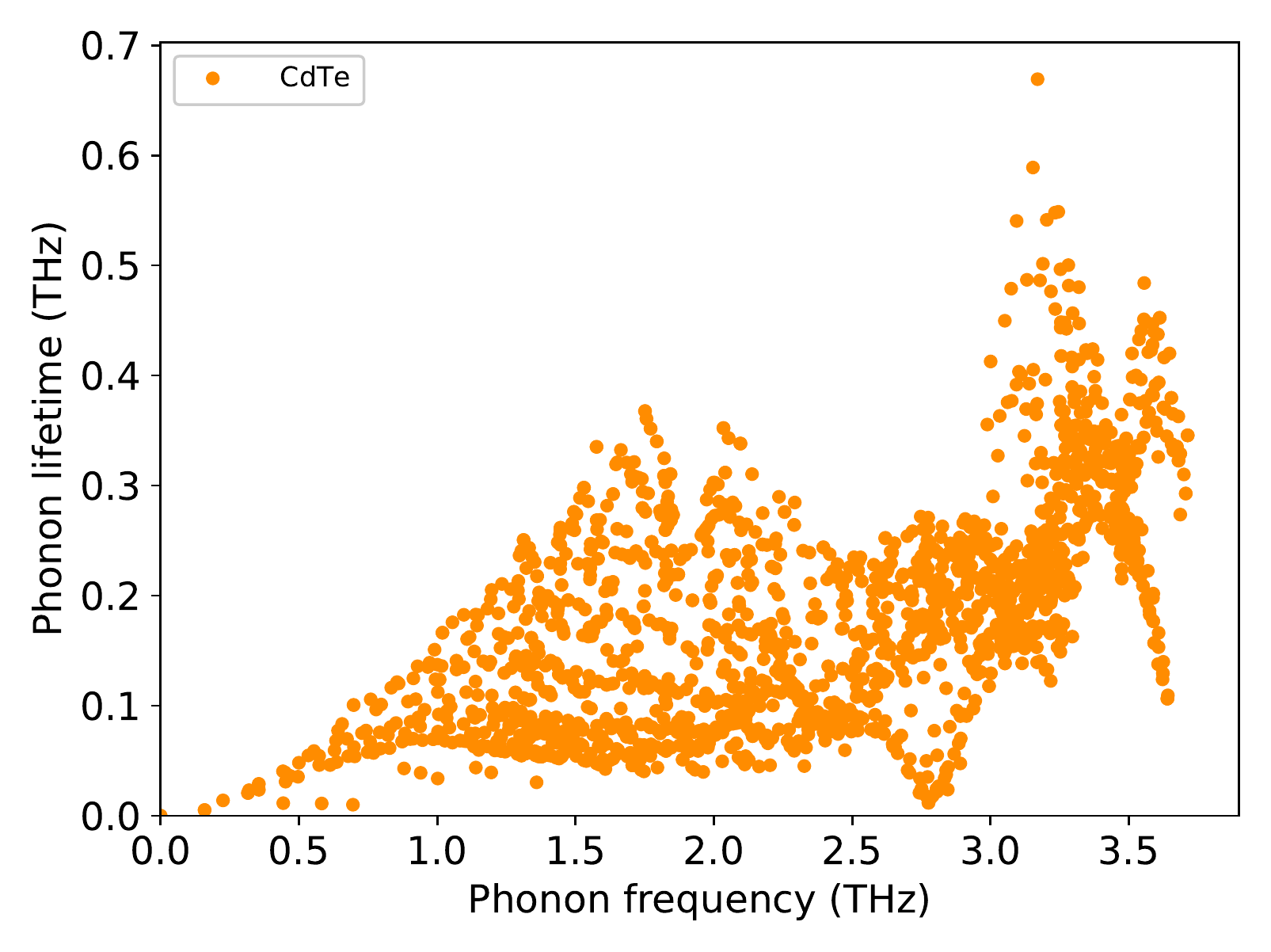}}
\mbox{\includegraphics[scale=0.35]{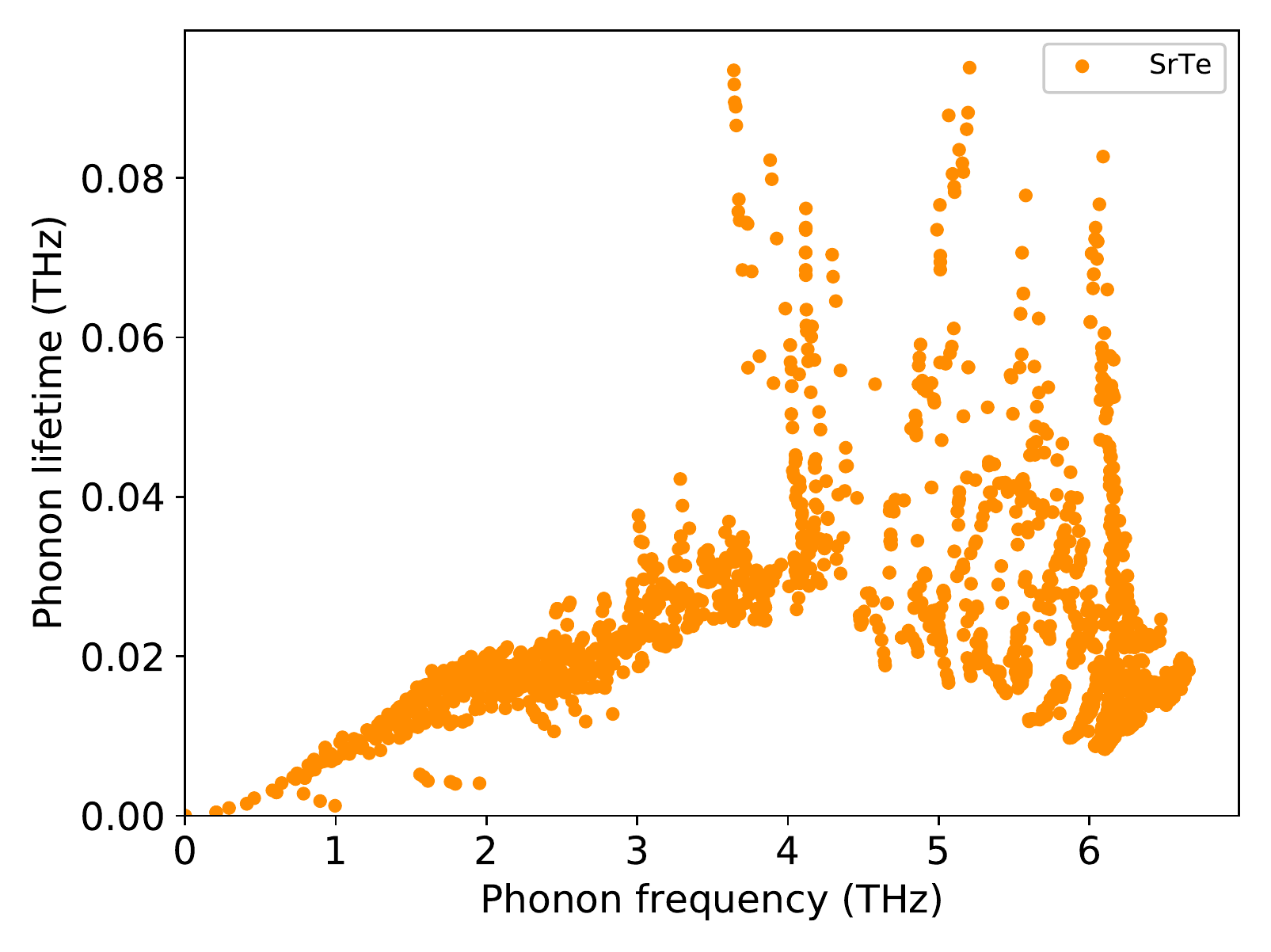}}
\mbox{\includegraphics[scale=0.35]{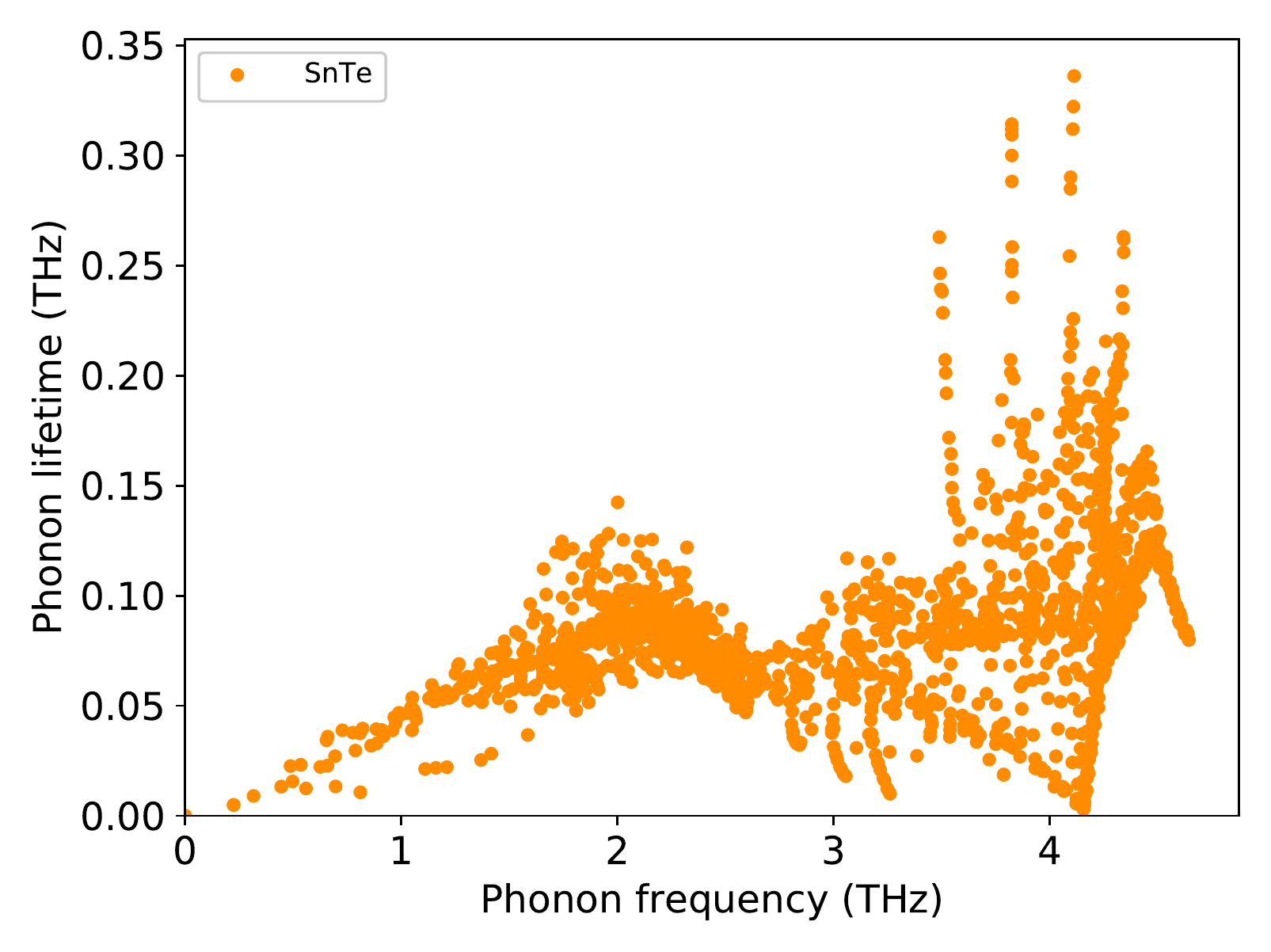}}
\end{subfigure}%
~
\begin{subfigure}[t]{0.5\textwidth}
\centering
\mbox{\includegraphics[scale=0.4]{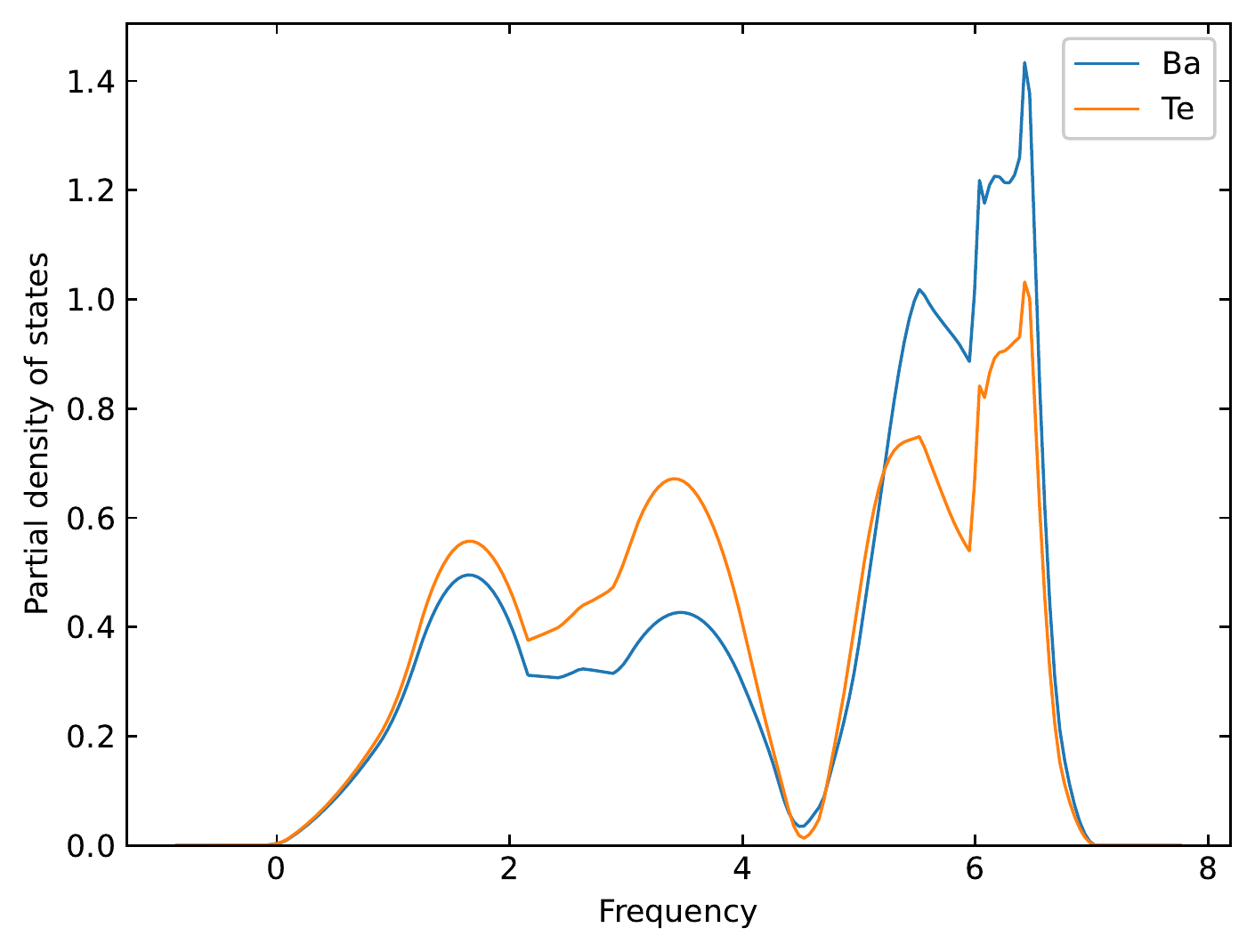}}
\mbox{\includegraphics[scale=0.4]{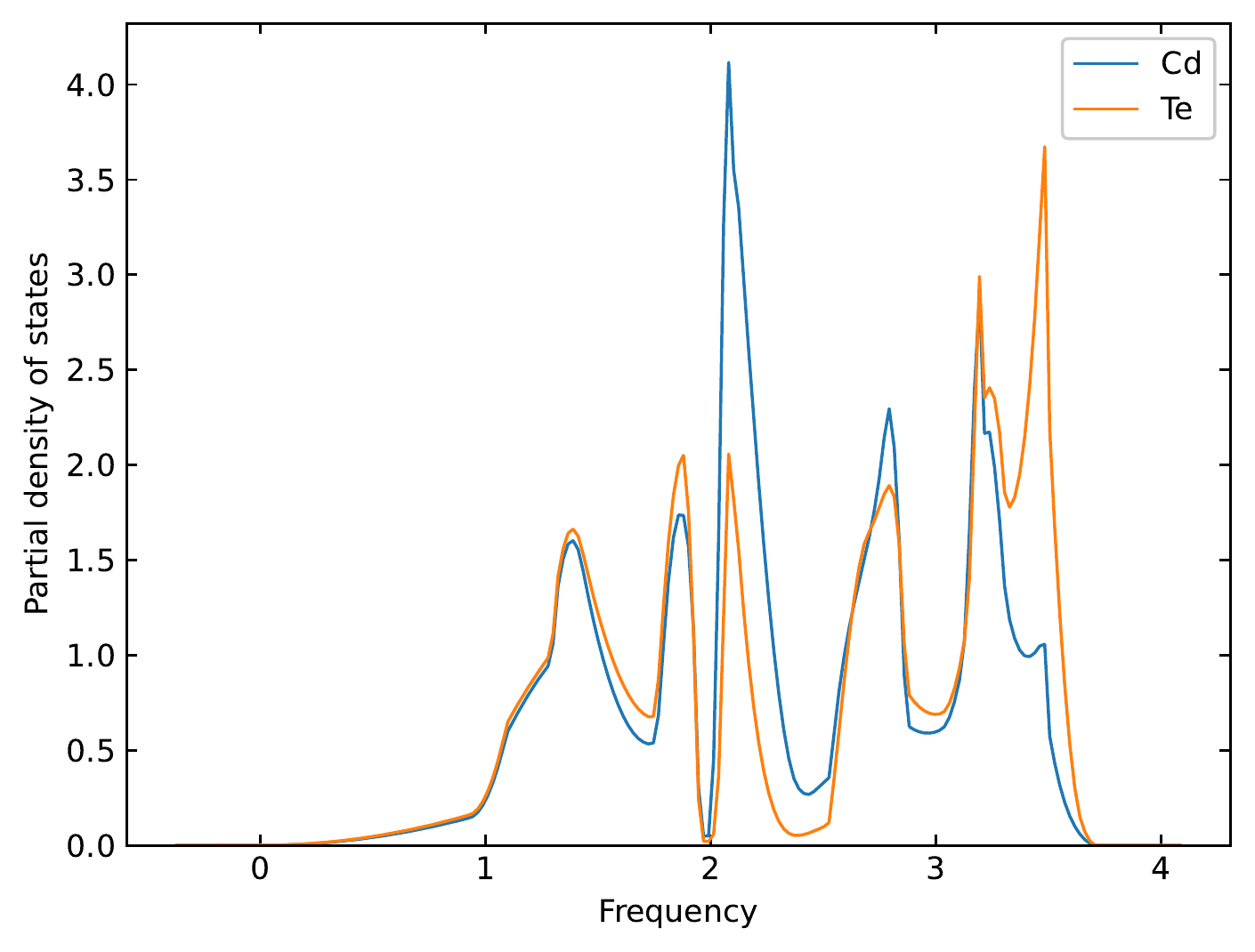}}
\mbox{\includegraphics[scale=0.4]{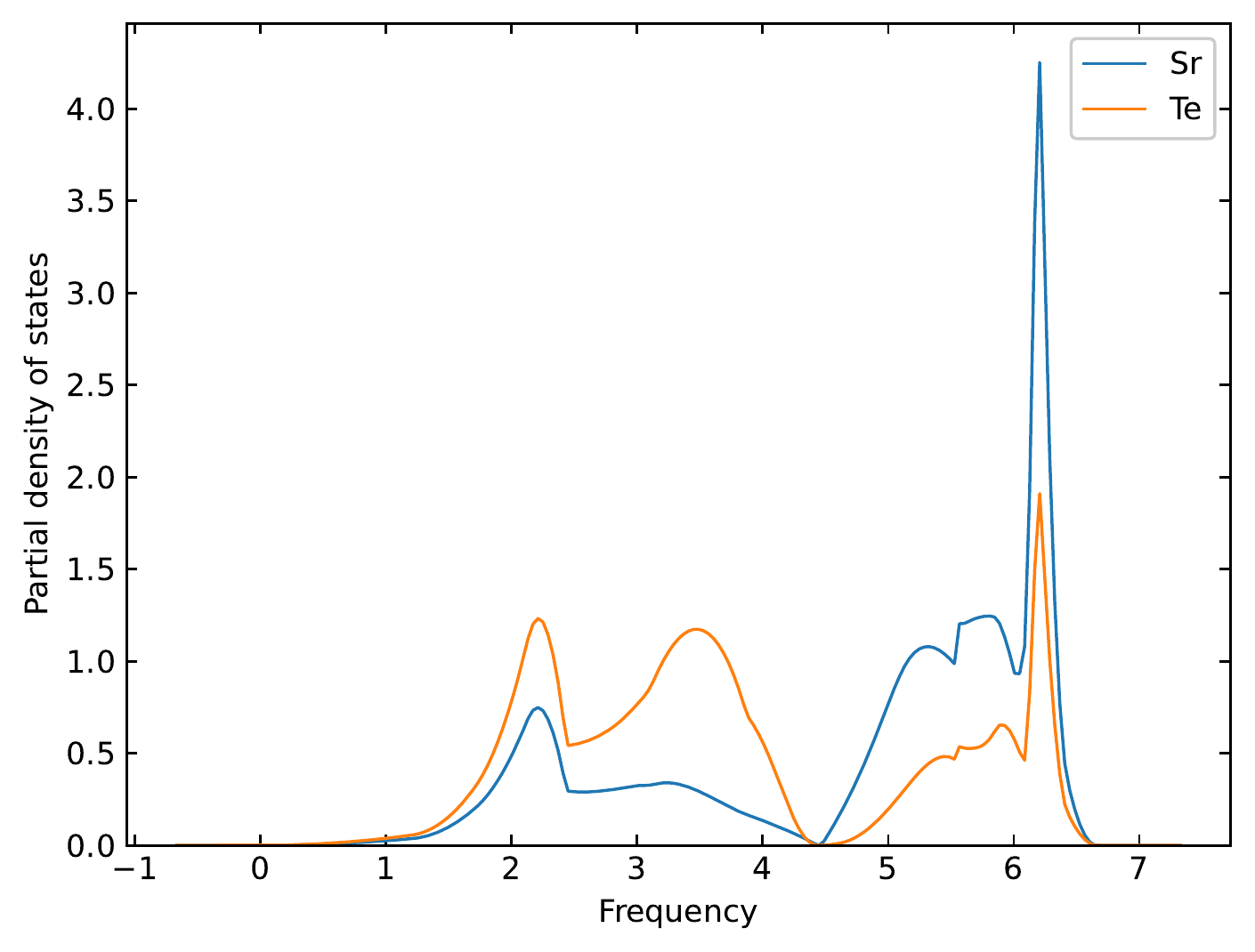}}
\mbox{\includegraphics[scale=0.4]{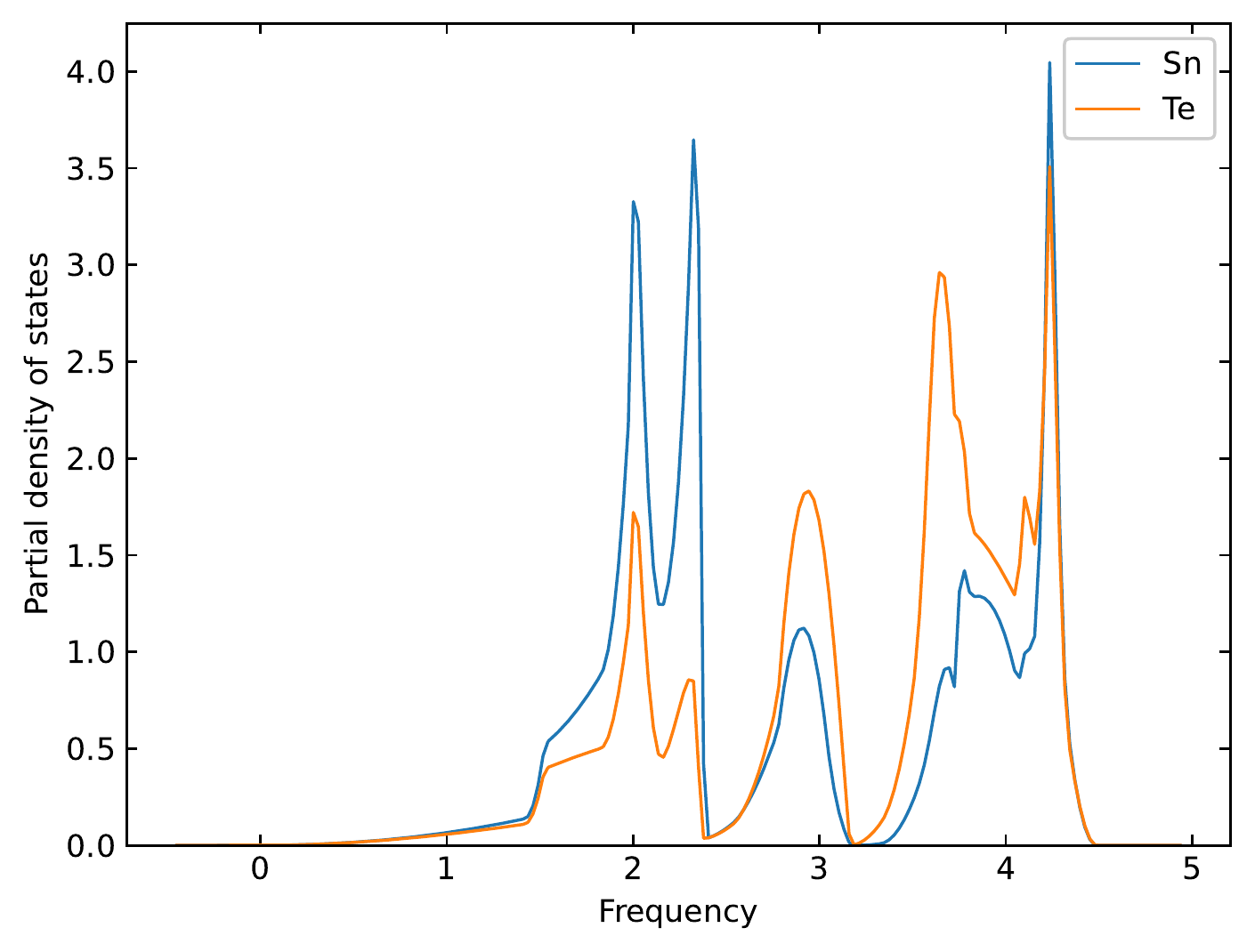}}
\end{subfigure}
\caption{\label{fig4} Phonon lifetime and Partial density of states of
a)~Phonon lifetime's scattered points~(left side).\ 
b)~Constituent contributor to total phonon density of States~(right side).\ 
Color online.\ Phonon lifetimes\ corresponds to respective participant element of partial mode density as a function of frequency of BaTe, CdTe, SrTe, and SnTe, respectively.}
\end{figure*}

%%%%%%%%%%%%%%%%%%%%%%%%%%%%%%%%%%%%%%%%%%%%%%%%%%%%%%%%%%%%%%
 Although each constituent atom contributes positive real eigenvalued phonon modes, overall density reveals that instable modes exist for unknown reasons in the entire system as viewed in Fig.~(\ref{fig3}). If it were a clean perturbed system, it would have the optimum optical gap depicted in Fig.~(\ref{fig2}).

%%%%%%%%%%%%%%%%%%%%%%%%%%%%Figure-3%%%%%%%%%%%%%%%%%%%%%%%%%%%%%
%%%%%%%%%%%%%%%%%%%%%%%%%%%%%%%%%%%%%%%%%%%%%%%%%%%%%%%%%%%%%%

 Despite the fact that Barium modes are more prevalent at lower frequencies and contribute higher proportion  to the total phonon life time at room temperature, however, as temperatures rise Ba and Te modes doing not much in thermal cooling processes. Tellurium and Cadmium each have their own specific point where they separate in lesser proportions, and boost the phonon life. When Strontium and Tellurium were alloyed, their trends looks the same, but the phonon life time contribution the list on the scale as compared to rest of systems. Tin alloyed with Tellurium, on the other hand, has a much better than SrTe; nevertheless, at higher frequencies, both Tin and Tellurium mutually contribute to the phonon life time, as seen in Fig.~(\ref{fig4}). 
%%%%%%%%%%%%%%%Figure 4%%%%%%%%%%%%%%%%%%%%%%%%%%%%%%

\begin{figure*}[htbp!]
\begin{subfigure}[t]{0.45\textwidth}
\centering
\mbox{\includegraphics[scale=0.35]{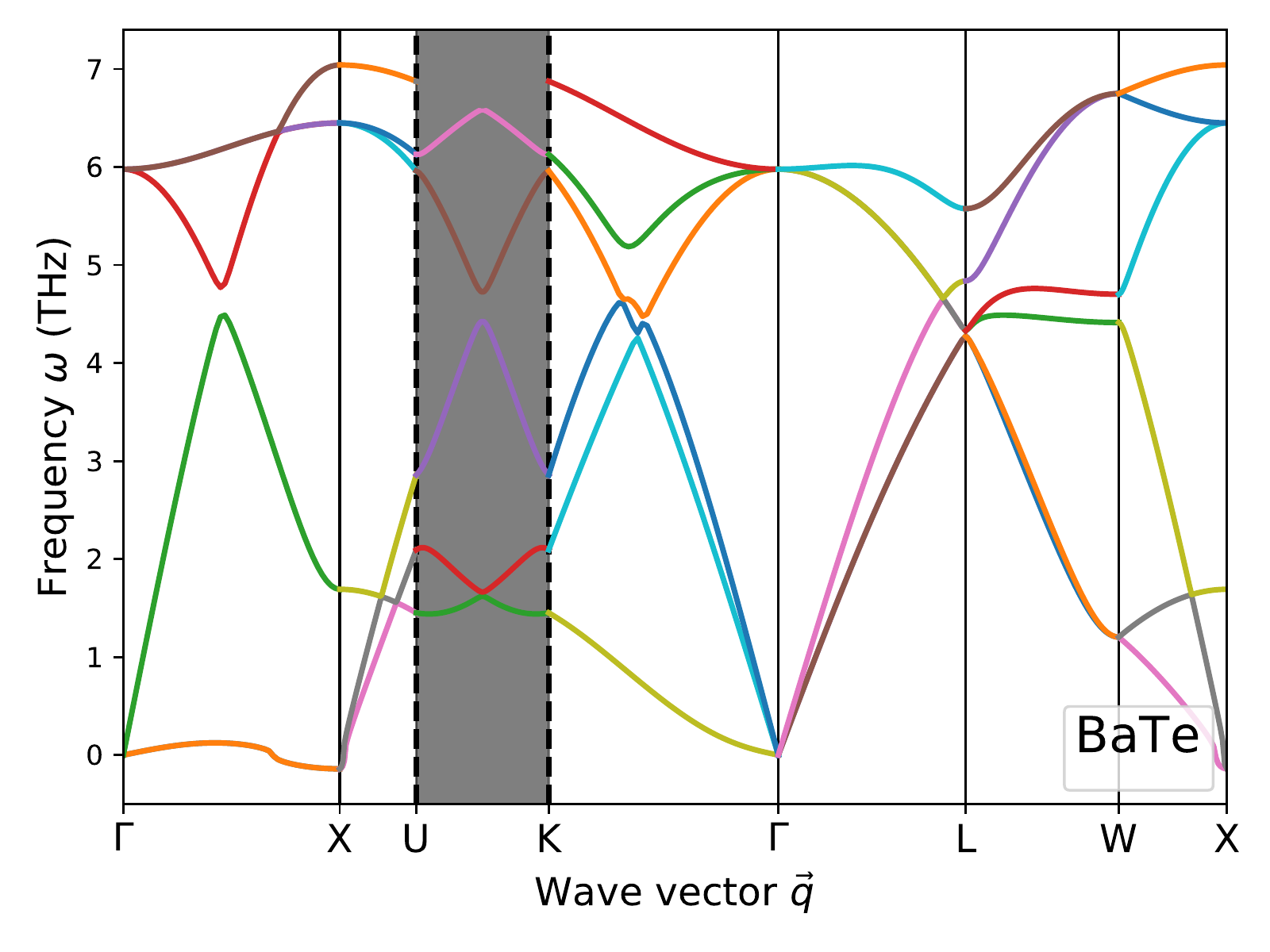}}
\mbox{\includegraphics[scale=0.35]{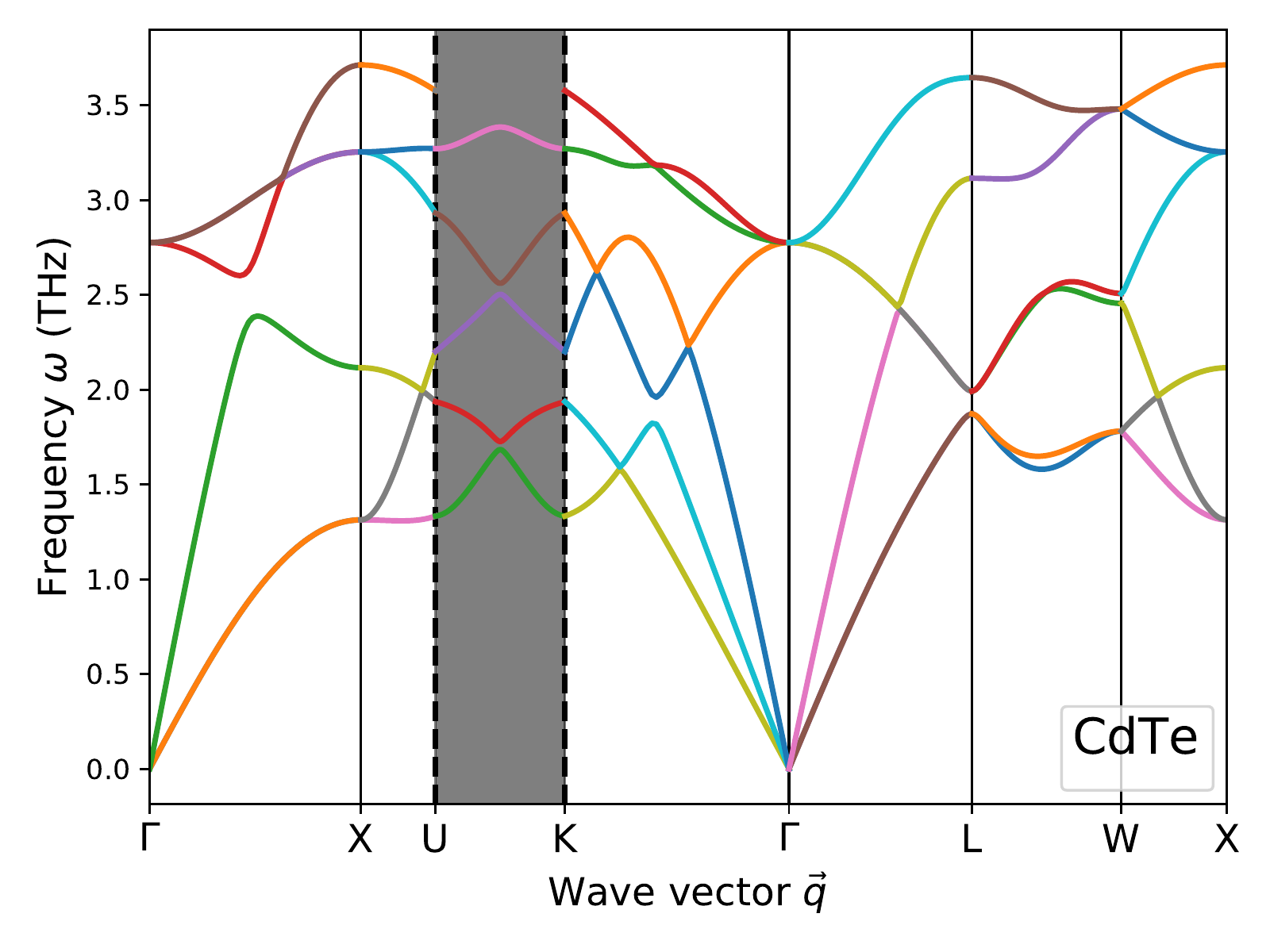}}
\mbox{\includegraphics[scale=0.35]{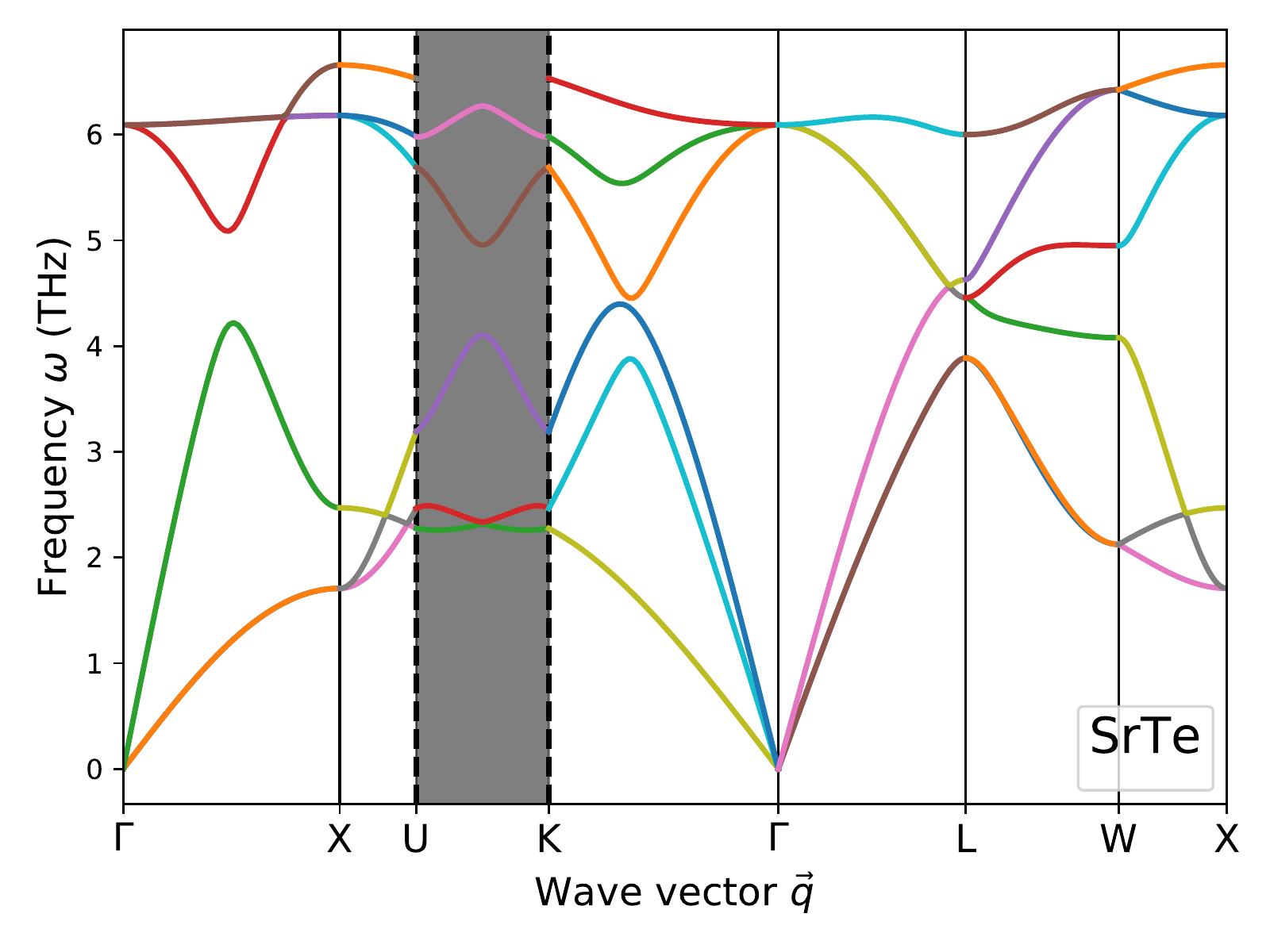}}
\mbox{\includegraphics[scale=0.35]{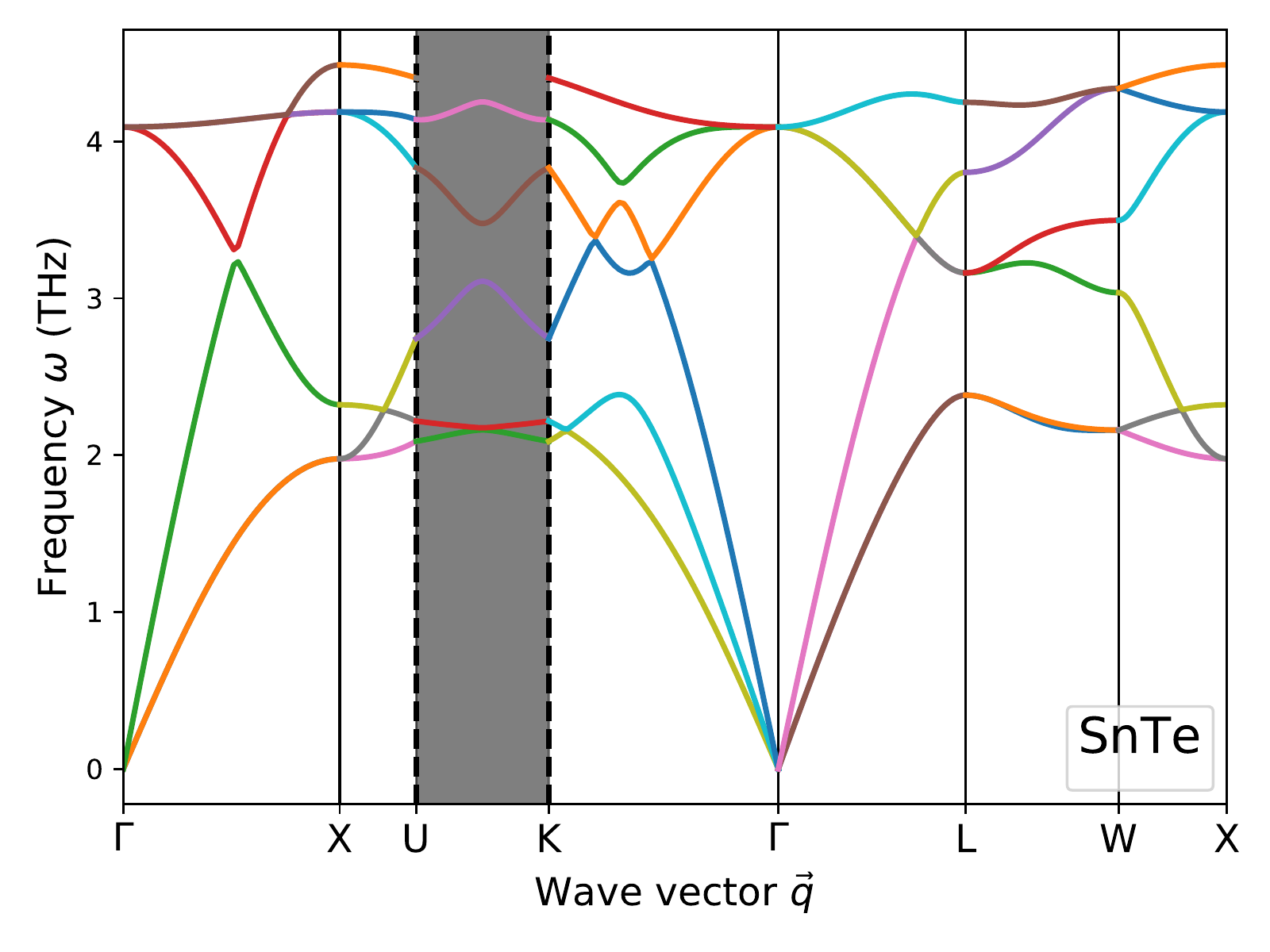}}
\end{subfigure}%
~
\begin{subfigure}[t]{0.45\textwidth}
\centering
\mbox{\includegraphics[scale=0.35]{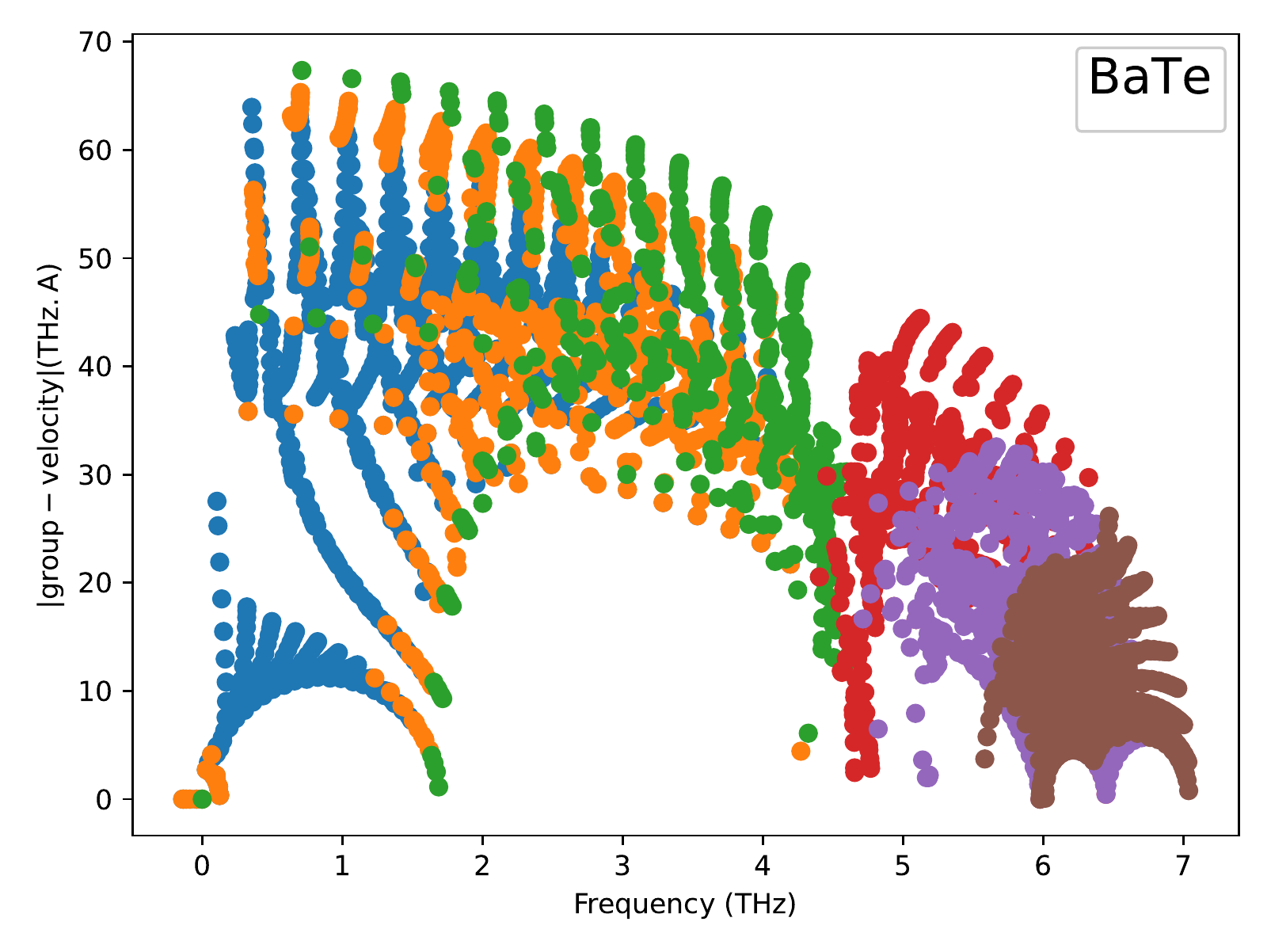}}
\mbox{\includegraphics[scale=0.35]{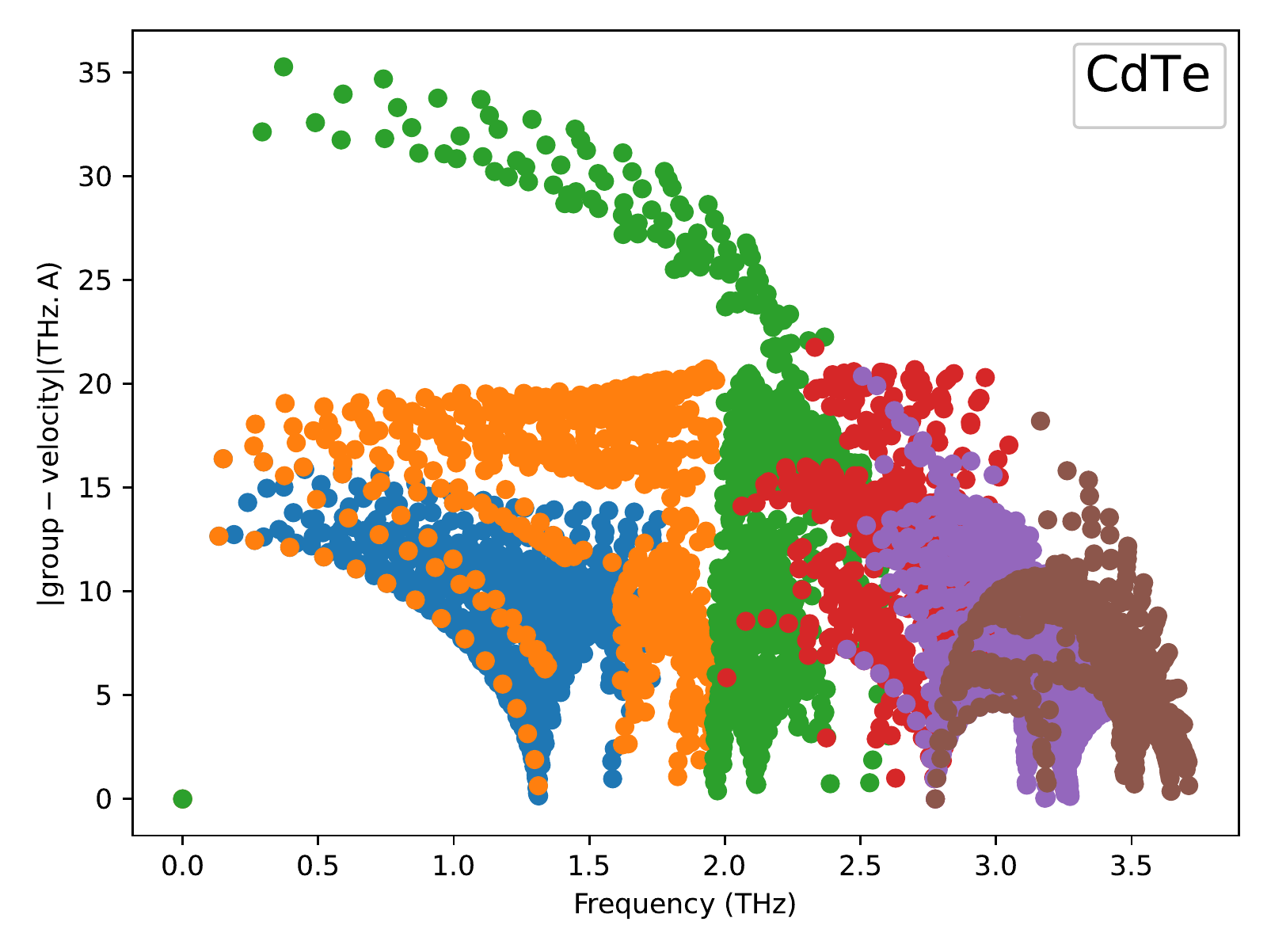}}
\mbox{\includegraphics[scale=0.35]{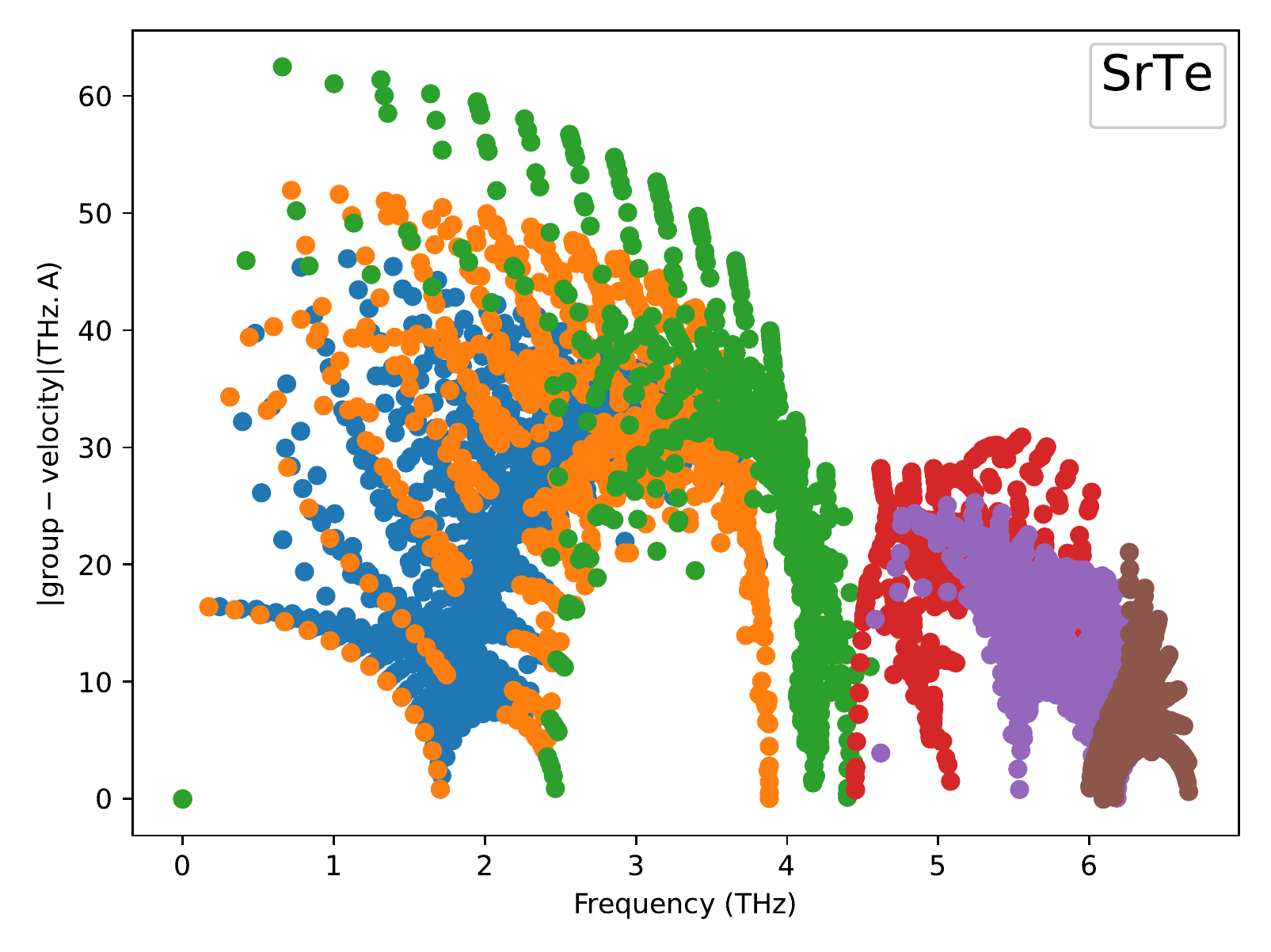}}
\mbox{\includegraphics[scale=0.35]{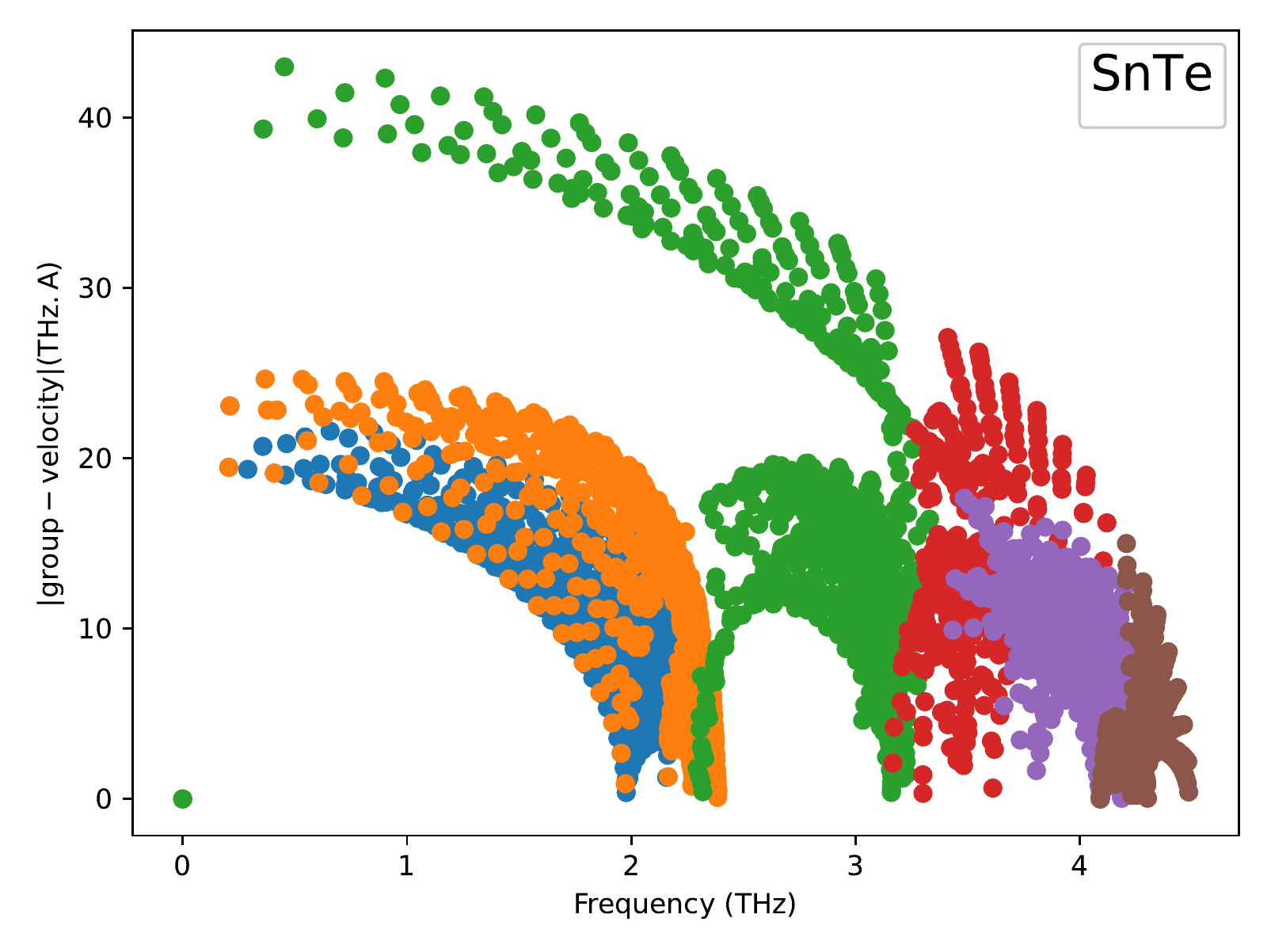}}
\end{subfigure}
\caption{\label{fig5} Phonon band structure and group velocities of
a)~Phonon dispersion~(left side).\ 
b)~Phonon group velocities~(right side).\ 
Color online.\ Phonon dispersion curves\ in all colors,\ corresponds to  similar color in group velocities as a function of frequency of different phonon modes for BaTe, CdTe, SrTe and SnTe, respectively.}
\end{figure*}

%%%%%%%%%%%%%%%%%%%%%%%%%%%%%%%%%%%%%%%%%%%%%%%%%%%%%%%%%%%%%%%%%%%%%%%%%%%%%%%%%%%5
Modes that oscillate in the same direction add up to optical modes in a spring model, but modes that oscillate against each other add up to acoustic modes. Surprisingly, as shown on left of Fig.~(\ref{fig5}), we have modes that execute the functions of optical and acoustic modes with improved cooling speed. Because the gap between amplitudon and electrical modes is narrow along the wave vector path, $\mathrm{K-\Gamma}$, BaTe, and SrTe display excellent electrical and thermal conductivity. CdTe and SnTe, on the other hand, have a broad gap and no sharp modulation amplitude, making both forms of conductivity difficult. A narrower gap has a higher group velocity than a wider gap, as illustrated on the right of Fig.~(\ref{fig5}).

%%%%%%%%%%%%%%%%%%%%%%%%Figure 5%%%%%%%%%%%%%%%%%%%%%%%

%%%%%%%%%%%%%%%%%%%%%%%%%%%%%%%%%%%%%%%%%%%%%%%%%%%%%%%%%%%%%%%%%%%%%%%%

\begin{figure*}[htbp!]
        \centering
        \includegraphics[scale=1.0]{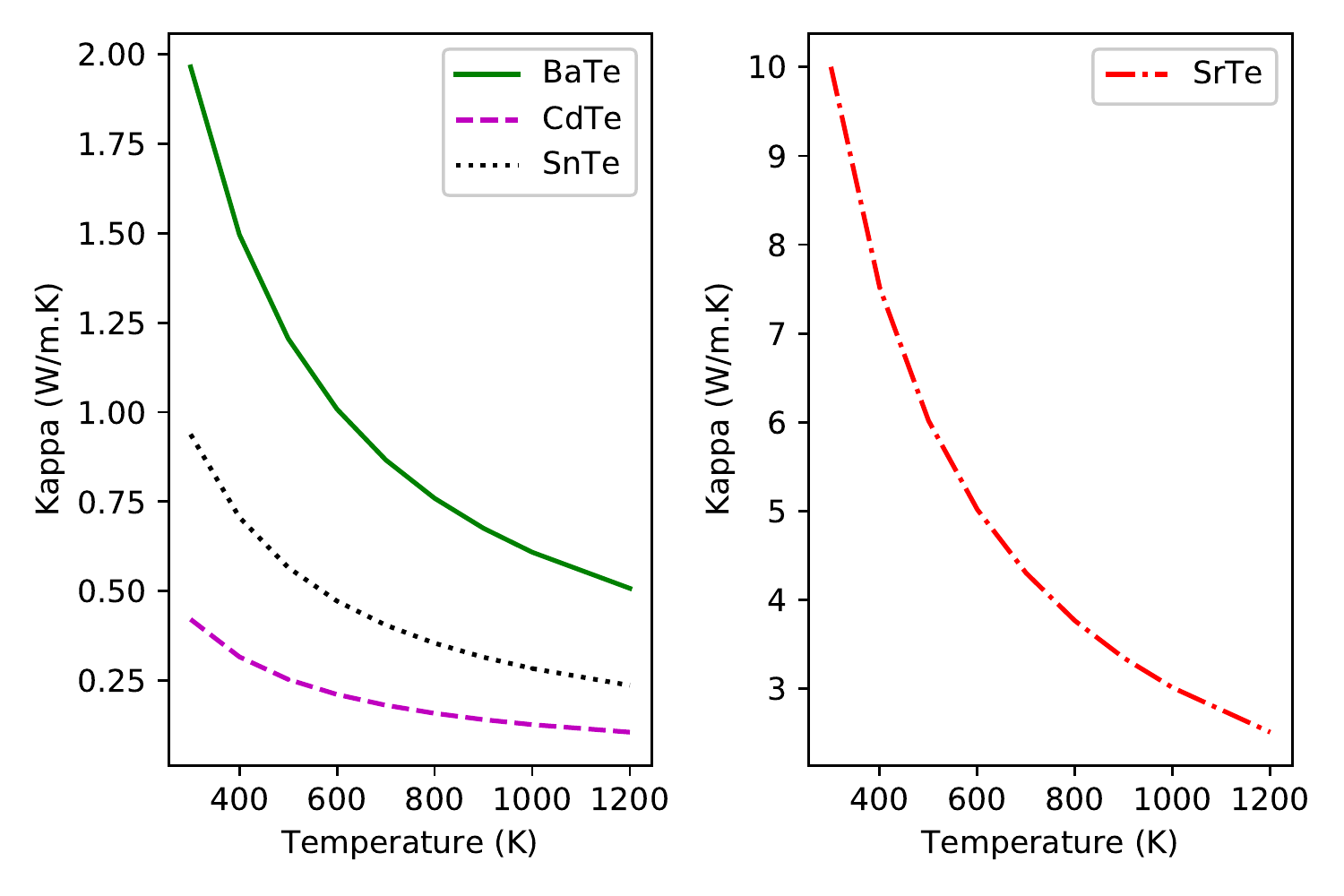}
        \caption{\label{fig6} Calculated lattice thermal conductivity
         of $\mathrm{\{Ba, Cd, \& Sr\}}$Te systems against temperature
         plot(Left), and relatively scaled SrTe dimmer on (Right).}
\end{figure*}

In thermoelectrics, thermal conductivity is the sum of electron-holes carrying heat ($\kappa_{e}$) and phonons passing through the lattice ($\kappa_{L}$). As a result, the thermoelectric figure of merit can be optimized by increasing electrical conductivity while decreasing thermal conductivity, alkaline earth tellurides are frequently employed as thermal conductivity moderators, increasing BaTe concentration reduces thermal conductivity of material in range [0, 2] as stated in Ref.~\cite{BKHJWGLS2011, LSHJBKKMDV2012}, which corresponds with our study as shown in Fig.~(\ref{fig6}).

The results in Ref.~\cite{KNNC1967} highlight experimentally determined pure CdTe thermal conductivity as 0.46~$\mathrm{W/m.K}$, which is identical to our result, 0.42~$\mathrm{W/m.K}$, and CdTe thermal conductivity shows a decrease over temperature range. Similarly, in Ref.~\cite{SHMZPYZWS2022}, the minimumal lattice thermal conductivity for SnTe is 0.81~$\mathrm{W/m.K}$, whereas our result is 0.82~$\mathrm{W/m.K}$. With the exception of SrTe, which has a thermal conductivity of 8.75~$\mathrm{W/m.K}$, comparably alkaline earth tellurides have a low thermal conductivity.

%%%%%%%%%%%%%%%%%%%%%%%%%%%%Figure-6%%%%%%%%%%%%%%%%%%%%%%%%%%%%%
%%%%%%%%%%%%%%%%%%%%%%%%%%%%%%%%%%%%%%%%%%%%%%%%%%%%%%%%%%%%%%

\section{Conclusion}

Here we investigate in-band localization in BaTe, CdTe, SrTe and SnTe. We studied several modes with group velocity ranging from sound speed to super sonic speed. Our findings explain the sharpening of longitudinal acoustic phonon modes.
It appears that heavier weighted Ba and Sr alloyed with tellurium have a much faster cooling nature due to sharp acoustic phonon modes. However, Cd and Sn seem to cool down just after a few nonlinear fluctuations. These findings demonstrate that it is possible to influence vibrational transport by doping heavy alkaline earth metal, which in turn allows for easier transportation of heat.

\section*{Disclosure\ statement}
The authors declare that there is no conflict of interest.

\section{Data\ Availability\ Statement}
The data that\ support the findings\ of\ 
this study\ are available\ upon reasonable\ 
request\ from the\ authors.\

\section{Acknowledgements}
We are grateful to the Ministry of Education 
of Ethiopia for~financial~support.~The authors 
also acknowledge~the International Science 
Program,~Uppsala~University,~Sweden,  for 
providing computer~facilities~for~research~at~the
Department~of~Physics.~The~office~of~VPRTT~of Addis
Ababa  university is also warmly~appreciated~for
supporting~this\ research under a grant~number~AR/053/2021.

\bibliography{sample}

\begin{thebibliography}{10}
\urlstyle{rm}
\expandafter\ifx\csname url\endcsname\relax
  \def\url#1{\texttt{#1}}\fi
\expandafter\ifx\csname urlprefix\endcsname\relax\def\urlprefix{URL }\fi
\expandafter\ifx\csname doiprefix\endcsname\relax\def\doiprefix{DOI: }\fi
\providecommand{\bibinfo}[2]{#2}
\providecommand{\eprint}[2][]{\url{#2}}

\bibitem{Poudel2008}
\bibinfo{author}{Poudel, B.} \emph{et~al.}
\newblock \bibinfo{journal}{\bibinfo{title}{High-thermoelectric performance of
  nanostructured bismuth antimony telluride bulk alloys}}.
\newblock {\emph{\JournalTitle{Science}}} \textbf{\bibinfo{volume}{320}}
  (\bibinfo{year}{2008}).

\bibitem{CWood1988}
\bibinfo{author}{Wood, C.}
\newblock \bibinfo{journal}{\bibinfo{title}{Materials for thermoelectric energy
  conversion}}.
\newblock {\emph{\JournalTitle{Reports on Progress in Physics}}}
  \textbf{\bibinfo{volume}{51}} (\bibinfo{year}{1988}).

\bibitem{XWTXYYZQT2009}
\bibinfo{author}{Xie, W.}, \bibinfo{author}{Tang, X.}, \bibinfo{author}{Yan,
  Y.}, \bibinfo{author}{Zhang, Q.} \& \bibinfo{author}{Tritt, T.~M.}
\newblock \bibinfo{journal}{\bibinfo{title}{Unique nanostructures and enhanced
  thermoelectric performance of melt-spun bisbte alloys}}.
\newblock {\emph{\JournalTitle{Applied Physics Letters}}}
  \textbf{\bibinfo{volume}{94}} (\bibinfo{year}{2009}).

\bibitem{NMTT1999}
\bibinfo{author}{Nolas, G.~S.}, \bibinfo{author}{Morelli, D.~T.} \&
  \bibinfo{author}{Tritt, T.~M.}
\newblock \bibinfo{journal}{\bibinfo{title}{Skutterudites: A
  phonon-glass-electron crystal approach to advanced thermoelectric energy
  conversion applications}}.
\newblock {\emph{\JournalTitle{Annual Review of Materials Science}}}
  \textbf{\bibinfo{volume}{29}} (\bibinfo{year}{1999}).

\bibitem{Delaire2011}
\bibinfo{author}{Delaire, J., O.and~Ma} \emph{et~al.}
\newblock \bibinfo{journal}{\bibinfo{title}{Giant anharmonic phonon scattering
  in pbte}}.
\newblock {\emph{\JournalTitle{Nature Materials}}}
  \textbf{\bibinfo{volume}{10}} (\bibinfo{year}{2011}).

\bibitem{Chen2017}
\bibinfo{author}{Chen, Z.} \emph{et~al.}
\newblock \bibinfo{journal}{\bibinfo{title}{Vacancy-induced dislocations within
  grains for high-performance pbse thermoelectrics}}.
\newblock {\emph{\JournalTitle{Nature Communications}}}
  \textbf{\bibinfo{volume}{8}} (\bibinfo{year}{2017}).

\bibitem{Manley2018}
\bibinfo{author}{Manley, M.~E.} \emph{et~al.}
\newblock \bibinfo{journal}{\bibinfo{title}{Supersonic propagation of lattice
  energy by phasons in fresnoite}}.
\newblock {\emph{\JournalTitle{Nature Communications}}}
  \textbf{\bibinfo{volume}{9}} (\bibinfo{year}{2018}).

\bibitem{BPVB1980}
\bibinfo{author}{Bak, P.} \& \bibinfo{author}{von Boehm, J.}
\newblock \bibinfo{journal}{\bibinfo{title}{Ising model with solitons, phasons,
  and the devil's staircase}}.
\newblock {\emph{\JournalTitle{Phys. Rev. B}}} \textbf{\bibinfo{volume}{21}}
  (\bibinfo{year}{1980}).

\bibitem{MQRC1983}
\bibinfo{author}{Quilichini, M.} \& \bibinfo{author}{Currat, R.}
\newblock \bibinfo{journal}{\bibinfo{title}{Neutron evidence for an overdamped
  phason branch in incommensurate k2seo4}}.
\newblock {\emph{\JournalTitle{Solid State Communications}}}
  \textbf{\bibinfo{volume}{48}} (\bibinfo{year}{1983}).

\bibitem{Bryan2020}
\bibinfo{author}{Bryan, L., Matthew S.and~Fu} \emph{et~al.}
\newblock \bibinfo{journal}{\bibinfo{title}{Nonlinear propagating modes beyond
  the phonons in fluorite-structured crystals}}.
\newblock {\emph{\JournalTitle{Communications Physics}}}
  \textbf{\bibinfo{volume}{3}} (\bibinfo{year}{2020}).

\bibitem{SNHOMA2017}
\bibinfo{author}{Shulumba, N.}, \bibinfo{author}{Hellman, O.} \&
  \bibinfo{author}{Minnich, A.~J.}
\newblock \bibinfo{journal}{\bibinfo{title}{Intrinsic localized mode and low
  thermal conductivity of pbse}}.
\newblock {\emph{\JournalTitle{Phys. Rev. B}}} \textbf{\bibinfo{volume}{95}}
  (\bibinfo{year}{2017}).

\bibitem{Manley2011}
\bibinfo{author}{Manley, M.~E.}, \bibinfo{author}{Abernathy, D.~L.},
  \bibinfo{author}{Agladze, N.~I.} \& \bibinfo{author}{Sievers, A.~J.}
\newblock \bibinfo{journal}{\bibinfo{title}{Symmetry-breaking dynamical pattern
  and localization observed in the equilibrium vibrational spectrum of nai}}.
\newblock {\emph{\JournalTitle{Scientific Reports}}}
  \textbf{\bibinfo{volume}{1}} (\bibinfo{year}{2011}).

\bibitem{MJBZA2014}
\bibinfo{author}{Manley, M.~E.} \emph{et~al.}
\newblock \bibinfo{journal}{\bibinfo{title}{Multiple high temperature
  transitions driven by dynamical structures in nai}}.
\newblock {\emph{\JournalTitle{Phys. Rev. B}}} \textbf{\bibinfo{volume}{89}}
  (\bibinfo{year}{2014}).

\bibitem{FCGRC2006}
\bibinfo{author}{Visoly-Fisher, I.}, \bibinfo{author}{Cohen, S.},
  \bibinfo{author}{Gartsman, K.}, \bibinfo{author}{Ruzin, A.} \&
  \bibinfo{author}{Cahen, D.}
\newblock \bibinfo{journal}{\bibinfo{title}{Understanding the beneficial role
  of grain boundaries in polycrystalline solar cells from single-grain-boundary
  scanning probe microscopy}}.
\newblock {\emph{\JournalTitle{Advanced Functional Materials}}}
  \textbf{\bibinfo{volume}{16}},
  \doiprefix\url{https://doi.org/10.1002/adfm.200500396}
  (\bibinfo{year}{2006}).

\bibitem{CFK2006}
\bibinfo{author}{Klingshirn, C.}
\newblock \emph{\bibinfo{title}{Optical Properties of Phonons}}
  (\bibinfo{publisher}{Springer Berlin Heidelberg}, \bibinfo{year}{2007}).

\bibitem{HWK2009}
\bibinfo{author}{Haug, H.} \& \bibinfo{author}{Koch, S.~W.}
\newblock \emph{\bibinfo{title}{Quantum Theory of the Optical and Electronic
  Properties of Semiconductors}} (\bibinfo{publisher}{WORLD SCIENTIFIC},
  \bibinfo{year}{2009}), \bibinfo{edition}{5th} edn.

\bibitem{TIT2015}
\bibinfo{author}{Togo, A.} \& \bibinfo{author}{Tanaka, I.}
\newblock \bibinfo{journal}{\bibinfo{title}{First principles phonon
  calculations in materials science}}.
\newblock {\emph{\JournalTitle{Scripta Materialia}}}
  \textbf{\bibinfo{volume}{108}} (\bibinfo{year}{2015}).

\bibitem{TACLTI2015}
\bibinfo{author}{Togo, A.}, \bibinfo{author}{Chaput, L.} \&
  \bibinfo{author}{Tanaka, I.}
\newblock \bibinfo{journal}{\bibinfo{title}{Distributions of phonon lifetimes
  in brillouin zones}}.
\newblock {\emph{\JournalTitle{Phys. Rev. B}}} \textbf{\bibinfo{volume}{91}}
  (\bibinfo{year}{2015}).

\bibitem{BJHAFZFEEFATKAKHEP2022}
\bibinfo{author}{Brorsson, J.}, \bibinfo{author}{Hashemi, A.},
  \bibinfo{author}{Fan, Z.} \& \bibinfo{author}{Fransson}.
\newblock \bibinfo{journal}{\bibinfo{title}{Efficient calculation of the
  lattice thermal conductivity by atomistic simulations with ab initio
  accuracy}}.
\newblock {\emph{\JournalTitle{Advanced Theory and Simulations}}}
  \textbf{\bibinfo{volume}{5}} (\bibinfo{year}{2022}).

\bibitem{Blochl94}
\bibinfo{author}{Bloch, P.}
\newblock \bibinfo{journal}{\bibinfo{title}{Projector augmented-wave method}}.
\newblock {\emph{\JournalTitle{Phys. Rev. B}}} \textbf{\bibinfo{volume}{50}}
  (\bibinfo{year}{1994}).

\bibitem{MHJ2005}
\bibinfo{author}{Mortensen, J.}, \bibinfo{author}{Hansen, L.} \&
  \bibinfo{author}{Jacobsen, K.}
\newblock \bibinfo{journal}{\bibinfo{title}{Real-space grid implementation of
  the projector augmented wave method}}.
\newblock {\emph{\JournalTitle{Phys.Rev.B}}} \textbf{\bibinfo{volume}{71}}
  (\bibinfo{year}{2005}).

\bibitem{KS65}
\bibinfo{author}{Kohn, W.} \& \bibinfo{author}{Sham, L.}
\newblock \bibinfo{journal}{\bibinfo{title}{Self-consistent equations including
  exchange and correlation effects}}.
\newblock {\emph{\JournalTitle{Phys. Rev.}}} \textbf{\bibinfo{volume}{140}}
  (\bibinfo{year}{1965}).

\bibitem{Enkovaaraetal2010}
\bibinfo{author}{Enkovaara, J.} \emph{et~al.}
\newblock \bibinfo{journal}{\bibinfo{title}{Electronic structure calculations
  with gpaw: a real-space implementation of the projector augmented-wave
  method}}.
\newblock {\emph{\JournalTitle{J. Phys. Condens. Matter}}}
  \textbf{\bibinfo{volume}{22}} (\bibinfo{year}{2010}).

\bibitem{VSYJCMFPH2021}
\bibinfo{author}{Shukla, V.}, \bibinfo{author}{Jiao, Y.},
  \bibinfo{author}{Frostenson, C.~M.} \& \bibinfo{author}{Hyldgaard, P.}
\newblock \bibinfo{journal}{\bibinfo{title}{vdw-df-ahcx: a range-separeted van
  der waals density functional hybrid}}.
\newblock {\emph{\JournalTitle{Journal of Physics: Condensed Matter}}}
  \textbf{\bibinfo{volume}{34}} (\bibinfo{year}{2021}).

\bibitem{EFFEEP2019}
\bibinfo{author}{Eriksson, F.}, \bibinfo{author}{Fransson, E.} \&
  \bibinfo{author}{Erhart, P.}
\newblock \bibinfo{journal}{\bibinfo{title}{The hiphive package for the
  extraction of high-order force constants by machine learning}}.
\newblock {\emph{\JournalTitle{Advanced Theory and Simulations}}}
  \textbf{\bibinfo{volume}{2}} (\bibinfo{year}{2019}).

\bibitem{CL2013}
\bibinfo{author}{Chaput, L.}
\newblock \bibinfo{journal}{\bibinfo{title}{Direct solution to the linearized
  phonon boltzmann equation}}.
\newblock {\emph{\JournalTitle{Phys. Rev. Lett.}}}
  \textbf{\bibinfo{volume}{110}} (\bibinfo{year}{2013}).

\bibitem{BKHJWGLS2011}
\bibinfo{author}{Biswas, K.} \emph{et~al.}
\newblock \bibinfo{journal}{\bibinfo{title}{High thermoelectric figure of merit
  in nanostructured p-type pbte–mte (m = ca{,} ba)}}.
\newblock {\emph{\JournalTitle{Energy Environ. Sci.}}}
  \textbf{\bibinfo{volume}{4}} (\bibinfo{year}{2011}).

\bibitem{LSHJBKKMDV2012}
\bibinfo{author}{Lo, S.-H.}, \bibinfo{author}{He, J.}, \bibinfo{author}{Biswas,
  K.}, \bibinfo{author}{Kanatzidis, M.~G.} \& \bibinfo{author}{Dravid, V.~P.}
\newblock \bibinfo{journal}{\bibinfo{title}{Phonon scattering and thermal
  conductivity in p-type nanostructured pbte-bate bulk thermoelectric
  materials}}.
\newblock {\emph{\JournalTitle{Advanced Functional Materials}}}
  \textbf{\bibinfo{volume}{22}} (\bibinfo{year}{2012}).

\bibitem{KNNC1967}
\bibinfo{author}{Kelemen, F.}, \bibinfo{author}{Néda, A.},
  \bibinfo{author}{Niculescu, D.} \& \bibinfo{author}{Cruceanu, E.}
\newblock \bibinfo{journal}{\bibinfo{title}{On the thermal conductivity of cdte
  and some cdte1−xsx and cdte1−xsex solid solutions}}.
\newblock {\emph{\JournalTitle{physica status solidi (b)}}}
  \textbf{\bibinfo{volume}{21}},
  \doiprefix\url{https://doi.org/10.1002/pssb.19670210213}
  (\bibinfo{year}{1967}).

\bibitem{SHMZPYZWS2022}
\bibinfo{author}{Su, H.} \emph{et~al.}
\newblock \bibinfo{journal}{\bibinfo{title}{Snte thermoelectric materials with
  low lattice thermal conductivity synthesized by a self-propagating method
  under a high-gravity field}}.
\newblock {\emph{\JournalTitle{Phys. Chem. Chem. Phys.}}}
  \textbf{\bibinfo{volume}{24}} (\bibinfo{year}{2022}).

\end{thebibliography}

\end{document}